\DocumentMetadata{}
\documentclass[acmsmall, screen]{acmart}
\AtBeginDocument{%
  \providecommand\BibTeX{{%
    \normalfont B\kern-0.5em{\scshape i\kern-0.25em b}\kern-0.8em\TeX}}}

\setcopyright{acmlicensed}
\copyrightyear{2025}
\acmYear{2025}
\acmDOI{XXXXXXX.XXXXXXX}

\acmConference[Conference acronym 'XX]{Make sure to enter the correct
  conference title from your rights confirmation emai}{June 03--05,
  2024}{Woodstock, NY}

\acmISBN{978-1-4503-XXXX-X/18/06}

\usepackage{url}            
\usepackage{booktabs}       
\usepackage{amsfonts}       
\usepackage{nicefrac}       
\usepackage{microtype}      

\usepackage{dblfloatfix}
\usepackage{graphicx}
\usepackage{float}
\usepackage{xspace}

\usepackage{amssymb}

\usepackage{algorithm}
\usepackage{algpseudocode}
\usepackage{amsmath}
\algrenewcommand\algorithmicrequire{\textbf{Input:}}
\algrenewcommand\algorithmicensure{\textbf{Output:}}
\usepackage{newfloat}
\usepackage{listings}
\usepackage{lineno}

\floatstyle{ruled}
\newfloat{listing}{tb}{lst}{}
\floatname{listing}{Listing}

\setcopyright{acmlicensed}
\acmJournal{TOSEM}

\usepackage{overpic}
\usepackage{xspace}
\usepackage{enumitem}
\usepackage{multirow}
\usepackage{mdframed}

\usepackage[utf8]{inputenc} 
\usepackage[T1]{fontenc}    
\usepackage{url}            
\usepackage{booktabs}       
\usepackage{amsfonts}       
\usepackage{nicefrac}       
\usepackage{microtype}      
\usepackage{xcolor}         
\usepackage{nameref}
\usepackage{soul}
\usepackage{color}
\usepackage{colortbl}

\newcommand{\argmax}{arg\,max}
\newcommand{\argmin}{arg\,min}

\definecolor{royalblue}{RGB}{65,105,225} 
\usepackage{xspace}
\makeatletter
\DeclareRobustCommand\onedot{\futurelet\@let@token\@onedot}
\def\@onedot{\ifx\@let@token.\else.\null\fi\xspace}
\def\eg{\emph{e.g}\onedot} 
\def\ie{\emph{i.e}\onedot}

\def\etal{\emph{et al}\onedot}

\makeatother

\definecolor{blue}{RGB}{0,102,204}  

\definecolor{lightgray}{HTML}{eeeeee}
\definecolor{darkgray}{HTML}{d8d8d8}
\definecolor{tablecellgreen}{HTML}{d0e6d0}

\definecolor{highlightColor}{rgb}{1, 0.8, 0.6}

\definecolor{amii_magenta}{HTML}{bf477c}
\definecolor{amii_summer}{HTML}{ffcccc}
\definecolor{amii_mustard}{HTML}{faa53c}
\definecolor{amii_sky}{HTML}{6c98ab}
\definecolor{amii_emerald}{HTML}{006c65}
\definecolor{amii_night}{HTML}{003f58}

\definecolor{top1Color}{HTML}{ff770f}
\definecolor{top2Color}{HTML}{01847f}
\definecolor{top3Color}{HTML}{ffecb3}
\definecolor{propColor}{HTML}{002FA7}
\definecolor{jiayang_todo}{HTML}{2FAF38}
\sethlcolor{highlightColor}

\newcommand{\toponetext}[1]{\textcolor{top1Color}{\textbf{#1}}}
\newcommand{\toptwotext}[1]{\textcolor{top2Color}{\textbf{#1}}}

\usepackage{subcaption}

\newif\ifdisplaycontent
\displaycontenttrue  

\def\ourmethod{AcTracer}
\def\WD{\textbf{\textit{WD}}}
\def\KS{\textbf{\textit{K-S}}}

\definecolor{color1}{HTML}{E5E5E1}
\definecolor{color2}{HTML}{E0DDEF}
\definecolor{color3}{HTML}{F3D7CA}
\definecolor{color4}{HTML}{E7E2B6}
\definecolor{color5}{HTML}{EAD6FA}

\def\randomselection{\textit{RandomSelection}}
\def\csssampling{\textit{CSSampling}}
\def\cessampling{\textit{CESampling}}
\def\pacesampling{\textit{PACESampling}}
\def\diffusesampling{\textit{DiffuseSampling}}
\def\rhcsampling{\textit{RHCSampling}}

\newcommand{\circled}[1]{{\large \textcircled{\footnotesize #1}}}

\newcommand{\answerYes}[1]{\textcolor{blue}{[Yes]}}
\newcommand{\answerNo}[1]{\textcolor{blue}{[No]}}
\newcommand{\answerNA}[1]{\textcolor{blue}{[NA]}}
\newcommand{\answerPartial}[1]{\textcolor{blue}{[Partial]}}

\newcounter{finding}
\newenvironment{finding}
{
    \refstepcounter{finding}
	\begin{mdframed}[
    	nobreak=true,
    	linecolor=black,
    	roundcorner=12pt,
    	backgroundcolor=gray!05,
    	linewidth=0.5pt,
    	leftmargin=1pt,
    	rightmargin=1pt,
        innerleftmargin=5pt,
        innerrightmargin=5pt,
    	topline=true,
    	bottomline=true,
    	skipabove=10pt
	]
    \textbf{Answer to RQ\arabic{finding}}: 
}
{
    \end{mdframed}
    \vspace{3pt}
}

\begin{document}
\title{AcTracer: Active Testing of Large Language Model via Multi-Stage Sampling}


\author{Yuheng Huang}
\affiliation{%
  \institution{The University of Tokyo}
  \country{Japan}
  }
\email{yuhenghuang42@g.ecc.u-tokyo.ac.jp}

\author{Jiayang Song}
\affiliation{%
  \institution{University of Alberta}
  \country{Canada}
  }
\email{jiayan13@ualberta.ca}

\author{Qiang Hu}
\affiliation{%
  \institution{Tianjin University}
  \country{China}
  }
\email{qianghu@tju.edu.cn}

\author{Felix Juefei-Xu}
\affiliation{%
  \institution{New York University}
  \country{USA}
  }
\email{juefei.xu@gmail.com}

\author{Lei Ma}
\affiliation{%
  \institution{The University of Tokyo, Japan, and University of Alberta}
  \country{Canada}
  }
\email{ma.lei@acm.org}

\renewcommand{\shortauthors}{Y. Huang, J. Song, Q. Hu, F. Xu, and L. Ma}

\begin{abstract}    
    Performance evaluation plays a crucial role in the development life cycle of large language models (LLMs). It estimates the model's capability, elucidates behavior characteristics, and facilitates the identification of potential issues and limitations, thereby guiding further improvement.
    Given that LLMs' diverse task-handling abilities stem from large volumes of training data, a comprehensive evaluation also necessitates abundant, well-annotated, and representative test data to assess LLM performance across various downstream tasks. However, the demand for high-quality test data often entails substantial time, computational resources, and manual efforts, sometimes causing the evaluation to be inefficient or impractical.
    To address these challenges, researchers propose active testing, which estimates the overall performance by selecting a subset of test data.
    Nevertheless, the existing active testing methods tend to be inefficient, even inapplicable, given the unique new challenges of LLMs (\eg, diverse task types, increased model complexity, and unavailability of training data).
    To mitigate such limitations and expedite the development cycle of LLMs, in this work, we introduce {\ourmethod}, an active testing framework tailored for LLMs that strategically selects a small subset of test data to achieve a more accurate performance estimation for LLMs.
    {\ourmethod} utilizes both internal and external information from LLMs to guide the test sampling process, reducing variance through a multi-stage pool-based active selection.
    Our experiment results demonstrate that {\ourmethod} 
    achieves state-of-the-art performance compared to existing methods across various tasks.

\end{abstract}



\maketitle

\section{Introduction}
\label{sec:introduction}

Evaluating Large Language Models (LLMs) is crucial for assessing their performance and identifying potential limitations, which provides an understanding of LLMs' capability and facilitates guiding directions for future improvements. 
{However, due to the intricate autoregressive nature of LLMs~\cite{brown2020language}, their diverse task-oriented capabilities, and the large volumes of training data, thorough evaluations are often impractical within the current LLM development life cycle. The high costs associated with LLM execution on large test datasets further complicate this process. Additionally, LLMs evolve rapidly, with different base models and fine-tuned versions producing significantly varied outputs, necessitating frequent re-evaluation. These factors make LLM evaluation both time- and resource-intensive.}


{When a complete evaluation is impractical, sampling offers an effective approach to obtain an approximate estimation using only a subset of the test data.}

Although random sampling is a commonly used baseline across various domains~\cite{donoho2006compressed, majumdar2017random, ozkan2018randomized}, the estimation accuracy can be further improved in an active sampling manner~\cite{kossen2021active, kossen2022active, fu2023estimating, guerriero2024deepsample}, namely \textit{active testing}. Specifically, given test data $D_{test}$, we iteratively choose a data point $d$ to sample and obtain its label $p$. At each step $t$, we collect and analyze an existing drawn set $\mathcal{T}_{t-1} = \{(d_1, p_1), \ldots (d_{t-1}, p_{t-1})\}$ to actively decide the next point to label,~$(d_t, p_t)$. Finally, the performance on $\mathcal{T}_{n}$ is used for the estimation given labeling budget $n$. {It’s important to note that, although similar, \textit{active testing} differs from \textit{active learning}. As Kossen et al.~\cite{kossen2021active} pointed out, using active learning can inevitably introduce bias into the testing scenario. While various active sampling solutions based on either internal~\cite{chen2020practical, guerriero2024deepsample} or external information~\cite{li2019boosting, feng2020deepgini, kossen2021active} have proven successful for classification-based DNNs, LLMs present new challenges for these methods:}

(1) \textbf{More Complex Mechanisms.}
{LLMs are Transformer-based, autoregressive, multi-task models with billions of parameters. Their structure and generative characteristics complicate both internal and external analysis. Internally, neuron activations can vary significantly across different tasks~\cite{zou2023representation, tang2024language}, and the large model size makes activation-based neuron coverage criteria~\cite{pei2017deepxplore}, which is often commonly used for capability estimation~\cite{wang2021prioritizing, feng2020deepgini}, computationally intensive and challenging for practical use. Externally, LLMs can generate thousands of tokens, creating a complex distribution of confidence scores for each response~\cite{kuhn2023semantic, huang2023look, xiong2024can}. In both scenarios, identifying reliable indicators to guide the active testing process is difficult.}

(2) \textbf{Inaccessibility to Prior Knowledge.} 
Given existing prior knowledge (\eg, labelled training data), it is possible to leverage supervised learning to learn the relationship between models' output and their performance, thereby directly obtaining an overall estimation for unlabeled ones. 
However, the situation becomes intricate in the context of LLMs. 
The training data of LLMs are usually inaccessible, and even with the training data, linking training loss to task performance is non-trivial. {While training-based test estimation remains useful and important when abundant labeled data is available, enabling meta-learning for performance estimation~\cite{fu2023estimating}, we argue that it is equally essential to develop entirely \textbf{training-free} methods to address the diverse scenarios in LLM development.}

(3) \textbf{High Complexity of Aggregated Benchmarks.}
{In response to the emergent capabilities of LLMs across diverse downstream tasks, current model evaluations are shifting focus from single-task assessments to aggregated multi-task benchmarks~\cite{zhong2023agieval, liu2023agentbench} with quality metrics from diverse perspectives~\cite{wang2023decodingtrust, sun2024trustllm}.}
Within these aggregated benchmarks, LLM behavior can vary significantly across different tasks, posing challenges for conducting unbiased sampling-based evaluations. 
Given these challenges, in this paper, we aim to investigate the following research question: 

\textit{Can we design an active testing framework for LLMs that is purely unsupervised (as a plug-and-play tool) to enable label-efficient evaluations?}

\begin{figure*}[t]
    \centering
    \includegraphics[width=1.0\linewidth]{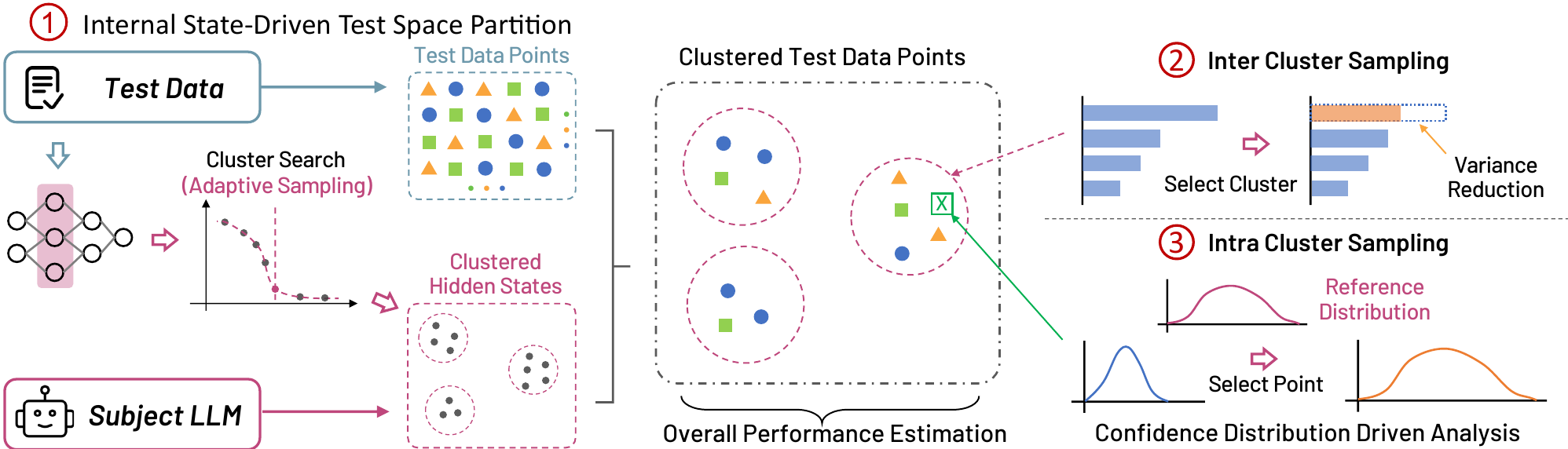}
    \caption{{Overall Workflow of {\ourmethod}.} 
    \circled{1}: {Internal State-Driven Test Space Partition involves the automatic, cluster-based partitioning of the test space, utilizing information collected from LLMs.}
    \circled{2}: An automated search mechanism to identify a suitable number of clusters corresponding to a given LLM and test data. The geometry structure obtained through the clustering algorithm partitions the input test spaces for more efficient sampling.  
    \circled{3}: An intra-cluster and inter-cluster sampling strategy to actively select the next data point for labeling.} 
    \label{fig:workflow}
\end{figure*}

We propose {\ourmethod}, {a testing framework} that leverages both internal (\eg, neuron activity) and external (\eg, output confidence score) information from LLMs to estimate the overall performance of the subject LLM in a pool-based multi-stage active selection manner. The overall workflow is illustrated in Fig~\ref{fig:workflow}. 
{The internal states of the models provide a unified representation across various tasks, naturally partitioning the entire test spaces into subpopulations where LLMs exhibit similar behavior patterns in each stratum.}
Then, the external confidence traces will be taken as more fine-grained indicators to guide the intra-cluster unbiased sampling.
{Extensive studies on seven datasets across different domains on five LLMs demonstrate the effectiveness of our method, which achieves state-of-the-art estimation accuracy for LLM evaluation.} Further ablation studies are conducted to investigate the impact of each component on our Framework.

{In summary, our primary contribution is the design and development of a generic and training-free active testing framework for LLMs. Additionally, we propose several automated methods to alleviate the burden of selecting hyperparameters within the extensive design spaces inherent to LLMs.}
We also implemented an extensive active testing testbed designed to support and enrich future research in this area. {We have included our accompanying source code as well as our experiment data at \url{https://github.com/YuhengHuang42/test_selection_llm}} 


\noindent \textbf{The Contributions to the Software Engineering Field.} {From a software engineering for artificial intelligence (SE4AI) perspective, LLM-driven intelligent software represents a new software paradigm that requires appropriate trustworthiness assurance methods for real-world applications. Among these methods, testing remains crucial for understanding system performance and robustness while providing guidance for future improvements. Our proposed method aims to enhance LLM testing workflow efficiency and substantially reduce associated costs through a practical, out-of-the-box, unsupervised approach. While efficient testing methods exist for classical Deep Neural Networks (DNNs)~\cite{li2019boosting, chen2020practical, guerriero2024deepsample}, LLMs present unique challenges. First, their billion-parameter architectures make many existing methods computationally impractical. Second, as foundation models, LLMs are deployed across diverse downstream tasks, necessitating universal testing approaches that can dynamically self-adjust. This paper represents one of the early attempts within the software engineering community to address these research gaps.}

{From the AI4SE perspective, this paper investigates the behavioral characteristics of LLMs in code generation tasks—one of their most significant applications in the software engineering domain today~\cite{hou2024large}. Our findings, observations, and insights aim to help stakeholders better understand LLM applications for software engineering tasks and illuminate future directions for improved testing, analysis, repair, and enhancement methodologies.}

\section{Background}
\label{sec:background}

In this section, we present the essential background, including an introduction to LLMs (Sec~\ref{sec:background:llms}), an overview of the current landscape in LLM evaluation (Sec~\ref{sec:background:evaluation}), and the problem setup for our study (Sec~\ref{sec:background:setup}).

\subsection{Large Language Models (LLMs)}
\label{sec:background:llms}

The underlying architecture of large language models (LLMs) is the Transformer~\cite{vaswani2017attention}, a widely recognized deep neural network (DNN) that has achieved remarkable success across various AI domains, including natural language processing~\cite{vaswani2017attention} and computer vision~\cite{dosovitskiy2021an}. LLMs differ fundamentally from earlier task-specific models in two key ways: 
(1): LLMs are pre-trained on massive datasets using unsupervised learning~\cite{radford2019language}, with a loss function typically defined as:

\begin{equation}
    \label{eq:next-token-pred}
    \small
    \max_{\theta} \sum_{t=1}^{T} \log p(w_t | w_{1:t-1}; \theta),
\end{equation}

where $w_t$ represents the token at time step $t$, and the model learns to predict tokens based on their context $w_{1:t-1}$ and updates its parameter $\theta$ accordingly.

(2) LLMs generate outputs in a token-by-token, autoregressive manner, striving to maximize the likelihood of the entire sequence, modeled as: 

\begin{equation}
    \label{eq:llm-inference}
    \small
    {\prod}_{t=1}^{T} p(\hat{w}_t = w_{t} | w_{1:t-1}; \theta), 
\end{equation}

where each token $\hat{w}_t$ is conditioned on all previous tokens $w_{1:t-1}$ in the sequence.

These two key mechanisms equip LLMs with unique capabilities: the pre-training stage enables them to understand and interpret knowledge across diverse domains, while the free-form autoregressive generation allows them to produce responses resembling those of a human expert. However, these mechanisms also introduce new challenges in model evaluation due to the vastly larger input and output spaces.

\subsection{LLM Evaluation}
\label{sec:background:evaluation}

Empirical evaluations of the human-like responses generated by LLMs are an ongoing hot topic in both research and industry communities. These evaluations focus on a wide range of quality aspects, including but not limited to factual consistency, faithfulness, robustness, toxicity, and hallucination on natural language processing tasks such as question-answering~\cite{lin-etal-2022-truthfulqa, wang2023decodingtrust, sun2024trustllm}, problem-solving skills~\cite{frieder2023mathematical} on mathematical problems~\cite{frieder2023mathematical}, and the efficiency, security, and correctness of generated code~\cite{evalplus, evalperf, sandoval2023lost}. 

Unlike the evaluation of models in the pre-LLM era, where output spaces were relatively fixed (e.g., classification tasks), assessing LLMs is significantly more complex due to their ability to generate thousands of tokens in a single response. This makes evaluation inherently non-trivial. {While some benchmarks provide ground truth references~\cite{rajpurkar-etal-2016-squad, yan-etal-2023-codetransocean}, simply measuring the similarity between reference and generated answers is often unreliable for both natural language and code. Correct answers can look completely syntactically dissimilar even if they are equally valid. Behavior testing offers an indirect way to assess generated outputs, but human evaluation remains the most reliable and transparent method for directly judging the quality of LLM outputs. However, relying on human evaluators increases labeling costs and can slow down continuous integration and deployment processes. }

{To address these challenges, there is growing interest in using LLMs themselves for evaluation~\cite{honovich-etal-2021-q2, santhanam2021rome, wang2024large}. Yet, recent studies show that LLM-based assessments can be biased and inconsistent~\cite{chen-etal-2024-humans, koo-etal-2024-benchmarking}. As a result, researchers and engineers still need to review these automated assessments to ensure their reliability. Additionally, running LLM evaluators requires significant computational resources.}

{These issues underscore the need for more efficient and unbiased evaluation frameworks for LLMs. A key scientific question is how to use as little labeled data as possible while still achieving accurate performance estimates. Traditionally, this involves sampling a small subset of data, labeling it, and using the results to estimate overall performance. However, there is potential to improve accuracy by incorporating additional feedback from the LLMs under evaluation.}


\subsection{Problem Setup}
\label{sec:background:setup}

The major goal of our study is to evaluate a specific property $P$  of a given LLM $f: \mathcal{X} \rightarrow \mathcal{Y}$ on its output space $\mathcal{Y}$. Here, $x = (x_1, x_2, \ldots, x_q) \in \mathcal{X}$ represents the input prompt with $q$ tokens to the LLM and $y = (y_1, y_2, \ldots, y_m) \in \mathcal{Y}$ denotes the generated response containing $m$ tokens. While the true value of the property $P$ is inaccessible, we can obtain an empirical estimation $\widetilde{P}$ of that given a test dataset $D_{test}$. Although collecting input-output pairs to form a large $D_{test}$ for accurate estimation is straightforward, evaluating the property (e.g., quality or trustworthiness) can be expensive and non-trivial and often requires experts' efforts or third-party API~\cite{lees2022new, markov2023holistic, zheng2023judging}. 


To reduce operational costs and streamline the CI/CD pipeline for LLMs, we propose selecting a small subset $\mathcal{T}_{n}$ for labeling within a budget of $n$ and using it to compute an accurate estimation of $\widetilde{P}$. To maximize the accuracy of this estimation, we adopt an active testing approach. In this method, the next data point $d_i = (x_i, y_i)$ to label at the $i$-th step is dynamically chosen based on the responses from the previously labeled $i-1$ points in $\mathcal{T}_{i-1}$.

\section{Methodology}
\label{sec:Methodology}
\subsection{General Framework}

The primary intuition behind AcTracer is that both internal and external information are important in guiding the testing of LLMs. Internal states can serve as a unified representation for LLMs' response to different tasks, which is crucial for a general framework; external output traces are important to categorize each individual test data point. 
Combining such information is crucial, as in the preliminary studies (shown in Section~\ref{sec:experiments:rq1}), we found that it is hard for either the internal-only method (e.g., \rhcsampling~\cite{guerriero2024deepsample}) or external-only method (e.g., \csssampling~\cite{li2019boosting}) to achieve robust and accurate estimations. Based on this intuition, {\ourmethod} encompasses the following three steps at a high level:

(1) Extract vector representations (internal neuron states) from LLMs for each data point in the test set and perform distance-based partitioning of the test set using these vectors;

(2) For each resulting cluster (stratum), actively select one for sampling based on online variance estimation, with the goal of minimizing overall estimation bias.

(3) Within each selected cluster, greedily choose a data point that minimizes the confidence distribution difference between the already selected data points and all the points in the cluster.

The overall algorithm is shown in Algorithm~\ref{algo:1}. 
For each stage, several key design settings significantly affect the estimation effectiveness (discussed in RQ2-RQ4). 
We present detailed explanations for the methodology behind {\ourmethod} in the following subsections.

\begin{algorithm}[H]
\caption{Overall Structure of {\ourmethod} }
\label{algo:1}
    \begin{algorithmic}[1]
        \Require LLM \(M\), Test input prompt set \(P\), Parameter \(\beta\), Sample budget \(n\), extracted features \(emb\)
        \Ensure Estimated performance \(\hat{\mu}\)
        \State \( k^* \leftarrow \text{Algo\_2}(M, P, emb) \) \Comment{Search for target cluster number $k^*$. Details in Appendix}
        \State \( C \leftarrow \text{Cluster}(M, P, k^*, emb) \) \Comment{Perform Balanced-K-means clustering.}
    
        \State \textbf{\# Initialize selection of each cluster with two samples following~\cite{carpentier2015adaptive} to avoid dividend by zero case.}
        \For{\(i = 1\) to \(k^*\)}
            \State \(S_i \gets \emptyset\)
            \For{\(m = 1\) to \(2\)}
                \State \textbf{\# Select the data point that minimizes the distribution distance (DIST) function.}
                \State \( \hat{q} \leftarrow \argmin_{q \in C_i \land q \notin S_i} \text{DIST}(C_i, S_i \cup \{q\}) \) 
                \State \( S_i \leftarrow S_i \cup \{\hat{q}\} \)
            \EndFor
        \EndFor
    
        \State \textbf{\# Begin Stratified Monte Carlo Sampling}
        \For{\( t = 2 \times k^* + 1\) to \(n\)}
            \For{\(k=1 \) to \(k^*\)}
                \State \( B_{k, t} \leftarrow \text{Compute\_B(\(C_{k}, t, \beta\))}\) \Comment{Compute B according to Eq.~\ref{eq:upper_confidence_bound}}
            \EndFor
            \State \( \hat{a} \leftarrow \argmax_{1 \leq a \leq k^*} B_{a, t} \)  \Comment{Select cluster to sample according to MC-UCB}
            \State \( \hat{q} \leftarrow \argmin_{q \in C_{\hat{a}} \land q \notin S_{\hat{a}}} \text{DIST}(C_{\hat{a}}, S_{\hat{a}} \cup \{q\}) \) \Comment{Intra-cluster sample}
            \State \( S_{\hat{a}} \leftarrow S_{\hat{a}} \cup \{\hat{q}\} \)
        \EndFor
    
        \State \( \hat{\mu} = \sum_{k=1}^{k^{*}} \frac{|C_{k}|}{|P|} \hat{\mu}_k \) \Comment{Compute the estimation given the mean of each cluster.}
        \State \Return \( \hat{\mu} \)
    \end{algorithmic}
\end{algorithm}

\subsection{Internal State-Driven Test Space Partition}
\label{sec:Methodology:clusearch}

Recent studies found that the internal states of LLMs contain valuable information capable of revealing important properties such as the LLMs' truthfulness~\cite{azaria2023the, chen2024inside}, knowledge~\cite{hendel-etal-2023-context, chen2024journey}, beliefs~\cite{hase-etal-2023-methods}, and emotions~\cite{zou2023representation}. 

Based on these findings, Our intuition ({also proved indirectly by our experiment results}) is that the internal neurons of LLMs often exhibit similar behavioral characteristics to human neurons, where different neurons exhibit diverse behavioral patterns in response to varying tasks~\cite{zarr2023foundations}. 
The hidden representation of neurons for each test point spans a high-dimensional space that represents the geometric structure of LLMs' internal reactions to presented queries. Within this space, test points associated with similar tasks tend to aggregate into compact groups.
As a result, the evaluation results for test points within the same subset are expected to be similar, which allows for a significant reduction in the number of tests required for accurate estimation.

Driven by this, the first step of our internal state-driven partition is to collect a series of hidden states by feeding the prompt to LLMs. 
Specifically, we draw the neural activities preceding the generation of the first token (\eg, LLMs' reactions to the prompt), which have been demonstrated to effectively represent the LLM's knowledge of the question~\cite{zou2023representation, ghandeharioun2024patchscope, hendel-etal-2023-context}. 
Ideally, all neuron activations within the LLM should be analyzed to form the representation of each data point. 
Nevertheless, given the computational constraints in real-world scenarios, particularly during the continuous development and integration phases of LLMs, such computation-intensive analysis is impractical. 
Therefore, we opt to take features from only one layer. Based on the findings from recent works, we select an intermediate layer (\eg, Layer 16 of a 32-layer LLM) as it has been recognized as the most informative for various downstream tasks~\cite{azaria2023the, zou2023representation, gurnee2024language, chen2024inside}. Additional results of utilizing the final layer for hidden state extraction are discussed in Section~\ref{sec:experiments:rq3}. To further reduce time complexity and avoid the curse of dimensionality~\cite{indyk1998approximate}, we apply Principal Component Analysis (PCA) for dimension reduction. 

With the extracted internal representations of each data point, unsupervised clustering algorithms can be applied to perform a distance-based partition on the vectorized test set. 
In this study, we select Balanced K-means~\cite{malinen2014balanced} as the partition algorithm, which is an adapted version of the original K-means that assigns an equal number of data points to each cluster. This algorithm aims to optimize the following the following target:

\begin{equation}\label{eq:bal_kmeans}
    \underset{a_1, \ldots, a_{D}}{\max} \sum_{d=1}^{D} -dist(\Vec{h}_{a_d}, \Vec{x}_{d}) \text{ s.t. } \forall k, \sum_{d=1}^{D} \mathbb{1}_{a_d =k} = \frac{D}{K}
\end{equation}

where $a_d \in \{0, \ldots, K\}$ is the cluster assignment index for each data point, $D$ is the number of test points, $dist$ is the distance function, $\Vec{h}_{a_d}$ is the cluster centers, $\Vec{x}_{d}$ is the hidden vector of $d$-th data point, $\mathbb{1}$ is the indicator function.

We choose this algorithm since (1) naive K-means can sometimes lead to extremely uneven partition sizes, which consequently lower the test estimation performance; 
(2) related work pointed out that Balanced K-means can achieve better performance for unsupervised domain discovery on LLMs~\cite{gururangan2023scaling}. Following ~\cite{gururangan2023scaling}, we use Equation~\ref{eq:bal_kmeans} for cluster center estimation and greedy inference when predicting clusters.

Given the candidate partition algorithm, the subsequent crucial step is to determine the number of clusters to optimize the partition process which significantly influences the clustering structure. To tackle this issue, we propose a solution called \textit{CluSearch} that performs an automated, model- and task-specific search to determine the best number of clusters. The designed search is empowered by the \textit{inertia} metric, namely, the objective function of naive K-means that measures the sum of distances between each data point and its corresponding cluster center:

\begin{equation}\label{eq:inertia}
    inertia = \sum_{{d} \in \{{0, \ldots, D}\}} min_{a_d \in \{0, \ldots, K\}} \|\Vec{x}_{d} - \Vec{h}_{a_d}\|^2
\end{equation}

In Eq. \ref{eq:inertia}, $D$ is the number of clusters, $\Vec{x}_{d}$ is the vector representation of each data point, $\Vec{h}_{a_d}$ is the cluster center. Given a fixed number of clusters, lower \textit{inertia} indicates better results. However, as this metric is a convex decreasing function in terms of cluster number, simply minimizing it by maximizing the number of clusters is trivial and ineffective. Instead, the relationship between cluster number and inertia is more of a trade-off, where the \textit{elbow} point of the cluster num-inertia curve is a widely used heuristic for appropriate cluster number search~\cite{thorndike1953belongs}. Mathematically, given a function $f$, the curvature of $f$ at point $x$ is:

\begin{equation}\label{eq:elbow}
    K_f(x) = \frac{f''(x)}{(1 + f'(x)^2)^{1.5}}
\end{equation}

where $f''(x)$ is the second derivative and $f'(x)$ is the first derivative. The elbow point is the point of maximum negative curvature of the curve. In this work, we utilized the Kneedle algorithm~\cite{satopaa2011finding} to find this point. To further enhance the efficiency of the search process, we leverage 
\textit{adaptive sampling}~\cite{nijholt2019adaptive} to intensively sample cluster number-inertia pairs in regions of rapid function change. This is achieved by iteratively dividing a given interval in the direction that can maximize the loss function:

\begin{equation}\label{eq:sample_loss}
    L_{lb, ub} = \sqrt{(ub - lb)^2 + (f(ub) - f(lb))^2}
\end{equation}

where $lb$ and $ub$ are the lower bound and upper bound of the interval, $f(x)$ is the inertia value at the point $x$.

In summary, our cluster number search algorithm is summarized in Algo.~\ref{algo:2}.

\begin{algorithm}
\caption{Search for Target Cluster Number}
\label{algo:2}
\begin{algorithmic}[1]
    \Require LLM \(M\), Test input prompt set \(P\), Target feature layer \(l^{*}\), Search budget \(w\), Search lower bound \(lb\), Search upper bound \(ub\)
    \Ensure Target cluster number at the elbow point
    \State \( SP \leftarrow \text{SPLIT}(lb, ub, w^{i}) \) \Comment{Split search interval into into $w^{i}$ equally spaced points} 
    \State \(ilist \gets \emptyset\) \Comment{Data structure for recording search history}
    
    \For{\(i = 0\) to \(w\)}
        \State \( cn \leftarrow \text{ADP\_SAMPLE}(ilist) \) \Comment{Perform \textit{adaptive} sampling given the past record}
        \State \(S_i \leftarrow \text{CLUSTER}(M, P, l^{*}, cn) \) \Comment{Perform Balanced-Kmeans given the cluster number $cn$ }
        \State \( ine_{i} \leftarrow \text{GET\_INERTIA}(S_i)\) \Comment{Compute inertial according to Eq.~\ref{eq:inertia}}
        \State \(ilist \leftarrow ilist \cup (cn, ine_{i}) \)
    \EndFor

    \State \(elb \leftarrow \text{FIND\_ELBOW}(ilist) \) \Comment{Perform Elbow-point detection based on search history}
    
    \State \Return \( elb \)
\end{algorithmic}
\end{algorithm}



\subsection{Inter-cluster Wise }

The partitioning achieved in the earlier phases naturally lends itself to the introduction of stratified sampling~\cite{GLASGOW2005683} for more efficient performance estimation. The primary goal of this strategy is to allocate the sampling budget across partitions to minimize the overall sampling error. If variances were known in advance, an optimal allocation strategy would be performable (\ie, distributing samples based on the variances within each cluster). 
However, this approach is impractical in our scenario, as the variance is not available during our pool-based active test data selection stage. 

To address this challenge, Carpentier \etal~\cite{carpentier2015adaptive} suggested an approach to progressively estimate variances. 
This method involves calculating the Monte Carlo Upper Confidence Bound (MC-UCB) for each cluster (treated as an ``arm'' in a multi-armed bandit problem) and selecting the arm with the highest upper bound for subsequent sampling. At current search round $t$, the MC-UCB score of cluster $k$ is computed as follows:

\begin{equation}\label{eq:upper_confidence_bound}
    B_{k,t} = \frac{w_{k}}{T_{k, t-1}}(\delta_{k, t-1} + \frac{2\beta}{\sqrt{T_{k, t-1}}}),
\end{equation}

where $w_k$ is the cluster size, $T_{k, t-1}$ is the number of points sampled in the previous round, ${\delta}_{k, t-1}$ is the empirical standard deviation within each cluster, and $\beta$ is a hyper-parameter. Under most LLM evaluation scenarios where the performance metric is bounded, the parameter $\beta$ can be set according to the number of sample $n$ as follows:

\begin{equation}\label{eq:beta}
    \beta = \sqrt{\log(2 / n^{-9/2})}
\end{equation}

Carpentier \etal provided formal proof of the convergence speed of the algorithm and the total MSE regret (i.e., the loss incurred by using the proposed algorithm instead of the optimal allocation) based on this parameter. The total regret is bounded by $poly(K)\widetilde{O}(n^{-7/6})$, where $K$ is the cluster number and $n$ is the number of data points. The algorithm assumes $n \leq 4K$ so we also impose this restriction when doing \textit{CluSearch} in Section~\ref{sec:Methodology:clusearch}

\subsection{Intra-cluster Wise}
\label{sec:Methodology:intra}
Although the inter-cluster sampling specifies the target cluster to apply sampling, it does not determine which specific data point in the cluster should get sampled and labeled. 
While random sampling remains a feasible option, more unbiased but resource-intensive sampling techniques can also be applied since the partition divides the space into smaller subsets, enabling high-complexity algorithms. 
Our intra-cluster sample is guided by the output \textit{confidence} of the LLMs. {While the internal states reflect models' knowledge,  the output confidence trace provides additional insights into the model’s behavior, which may vary depending on the decoding algorithms used.} 

{However, the effectiveness of this information largely depends on the aggregation method, namely, how the confidence scores for individual tokens are combined into a single score. 
For instance, when evaluating Llama-2 on TruthfulQA, we observed that the confidence scores of the first token from the generation are all below 0.006, with a standard deviation of approximately 8e-4. 
In contrast, averaging the conference scores across all tokens yields a score of 0.732 with a standard deviation of 0.076. The confidence distribution in the first token case is overly monolithic compared with the overall distribution of all generated tokens, making it difficult to differentiate between test points. 
Consequently, a poor choice of confidence aggregation could cause a confidence distribution-driven sampling method to degrade into random sampling.}

To address this challenge, we propose to select an aggregation method by first transforming the obtained confidence distribution into a histogram representation and calculating the entropy of that distribution. In this scenario, higher entropy indicates a more separated distribution over the test space, making it more informative from an information theory perspective. We conducted a dataset-level entropy analysis for four different aggregation methods: \textit{first} (models' confidence to the input prompt), \textit{max} (maximum confidence over the entire generation sequence), \textit{mean} (average confidence), and \textit{gmean} (geometric mean, which may better capture the autoregressive generation process). 
We leave more complicated aggregation methods for future work. 

After obtaining confidence scores for each data point, our goal in this intra-cluster stage is to maintain the confidence distribution of the sample drawn to be as close as possible to the distribution of the entire cluster, aiming for an intra-cluster level unbiased sampling.
This is achieved by selecting candidate sample points that greedily minimize the distance between the confidence distributions of the sampled points and the entire cluster. 
For measuring the distance between distributions, the two-sample Kolmogorov-Smirnov test (K-S)~\cite{pratt1981kolmogorov} and the Wasserstein Distance (\WD)~\cite{fournier2015rate, arjovsky2017wasserstein} are common choices. The {\KS} is more suitable for unimodal distributions, whereas the {\WD} is a better fit for assessing multimodal distributions~\cite{reschenhofer1997generalization, lee2019hierarchical}. 
Therefore, we conduct the dip test~\cite{hartigan1985dip} to assess the distribution's modality and decide which distance metric to use accordingly. This selection can bring, at most, a 61.38\% accuracy improvement when evaluating Code Llama on the HumanEval dataset. 


\section{Evaluation Design}
\label{sec:design}

In this section, we briefly describe the rationale behind our selection of the experiment dataset, metric, and baselines. Additionally, we detail the configurations used in our experiments for clarity and reproducibility.

\subsection{Dataset}

A key advancement of LLMs over their predecessors is their ability to handle diverse tasks. 
Taking this unique characteristic into account, we select seven evaluation datasets in eight settings to cover a range of model capabilities, including common knowledge, mathematical reasoning, problem-solving, and code generation. 
The included datasets are listed below:

\begin{itemize}[noitemsep, topsep=0pt, parsep=3pt, partopsep=0pt, leftmargin=*]
    \item \textbf{Common Knowledge.}
    We select TriviaQA \cite{joshi-etal-2017-triviaqa} and NQ-open \cite{lee-etal-2019-latent} as two question-answering datasets designed to evaluate the World Knowledge of LLM.
    \item \textbf{Mathematical Reasoning.}
    We evaluate our method on GSM8K \cite{cobbe2021gsm8k}, a dataset that focuses on basic math problems that require multi-step reasoning.
    \item \textbf{Problem Solving.}
    We use AGIEval \cite{zhong2023agieval}, an aggregated benchmark aimed at assessing LLMs within the context of human-centric standardized exams to gauge their general abilities in human cognition and problem-solving. 
    \item \textbf{Truthfulness.}
    We choose TruthfulQA \cite{lin-etal-2022-truthfulqa}, a benchmark tailored for imitative falsehoods measurement to assess the truthfulness and informativeness of the LLMs. In our experiment, we refer to informativeness evaluation as TruthfulQA-I and truthfulness evaluation as TruhtulQA-T.
    \item \textbf{Code Generation.}
    MBPP \cite{austin2021program} and HumanEval \cite{chen2021evaluating} datasets are selected to test LLMs' ability to understand human intentions and perform corresponding code generation.
\end{itemize}

We aim to select a range of diverse and representative datasets to perform a rigorous evaluation for the testing methods under complex conditions that closely mirror real-world scenarios. The datasets vary in size, from over one hundred (HumanEval) to nearly 18,000 (TriviaQA), spanning three orders of magnitude, as shown in Table~\ref{tab:datasize}. 
With the evaluation of various downstream tasks, we aim to reveal the strengths and weaknesses of different methods and offer practical guidelines for the following applications.

\begin{table*}[htbp]
    \centering
    \begin{tabular}{ccccccc}
    \toprule
        AGIEval & TriviaQA & NQ-open & GSM8K & TruthfulQA & MBPP & HumanEval \\ \midrule
        3852 & 17944 & 3610 & 1319 & 817 & 500 & 164 \\ \bottomrule
    \end{tabular}
    \caption{Number of data points for each dataset in our evaluation.} \label{tab:datasize}
    \vspace{-25pt}
\end{table*}

\subsection{Metric}

One common widely adopted approach to evaluate the effectiveness of active testing methods is to measure the errors between the estimation and the ground truth, typically using metrics like RMSE~\cite{guerriero2024deepsample}. 
However, in our pool-based setting, where data points are actively and progressively selected, a single-point estimation error may not provide a complete picture for the effectiveness assessment. 
To tackle this issue, in this study, we conduct evaluations for sampling proportion~(labeling budgets) $p$ ranging from 5\% to 50\% of the original dataset, with increments of 3\%, and use the results to construct a 2-D diagram. 
This diagram plots the number of sampling points (x-axis) against the relative error of the estimation (y-axis). 
We then calculate the Area Under the Curve (AUC) as an indicator of each method's effectiveness using $\int_t y(t) dt =\int_t y(t) \left.\frac{dx}{dt}\right|_{x=x(t)} dt$, where 
a lower AUC value indicates a better performance. 

\subsection{Baselines}
For the selection of baseline methods, we adhere to the following criteria: 
(1) the methods should function as plug-and-play tools for test estimation without the need for training data; 
(2) the methods selected should either be widely accepted in the industry, published in top-tier conferences or journals and proved to be useful for classical DNNs, or available as pre-print versions that demonstrate promising results on more recent LLMs. 
Aligned with these criteria, we selected five baseline methods:

\begin{itemize}[noitemsep, topsep=0pt, parsep=3pt, partopsep=0pt, leftmargin=*]
\item \csssampling~\cite{li2019boosting}, which enhances test estimation efficiency by dividing the models' confidence scores across the entire dataset into $k$ sections and then applying stratified sampling according to the confidence distribution within each bin.
\item \cessampling~\cite{li2019boosting}~\footnote{We optimized the code by implementing vectorization, achieving an approximate 100x speedup, thereby making the algorithm practical for LLMs.}, which guides the sampling process through distribution analysis of the last layers' neurons between selected points and the entire test set; 
\item \pacesampling~\cite{chen2020practical}, which utilizes the MMD-critic algorithm to select the most representative test inputs for test estimation.
\item \diffusesampling~\cite{ashury2024label}, a recent approach for label-efficient model selection of LLMs based on clustering analysis of their output text embeddings~\footnote{Model selection and performance estimation share similar settings. By assigning the ground truth label to LLMs' performance, this approach can be effectively adapted for test estimation.}. 
\item \rhcsampling~\cite{guerriero2024deepsample}, Rao, Hartley, and Cochran Estimator-based Sampling guided by Likelihood-based Surprise Adequacy~\cite{kim2019guiding} of internal neurons.
\end{itemize}

\subsection{Experiment Configurations} 

For our experiments, we selected Llama 2-7B~\cite{touvron2023llama} for natural language processing tasks and Code Llama-7B-Python~\cite{roziere2023code} for code generation tasks (e.g., HumanEval and MBPP) in the first set of experiments. {In the second set of experiments, we evaluated Llama-3-8B~\cite{dubey2024llama}, Qwen2.5~\cite{yang2024qwen2}, and Phi-4~\cite{abdin2024phi} on all settings because, as more advanced models, they also demonstrate moderate performance in code-related tasks.}
To mitigate the inherent randomness in the experiment, we repeat our experiments ten times and use the median relative error for the AUC computation. {All the experiments take at least {12512} CPU Hours and {1,588} GPU Hours on a server with AMD 3955WX CPU (3.9GHz), 256GB RAM, and four NVIDIA A4000 GPUs (16GB VRAM of each).



\section{Experiments and Analysis}
\label{sec:experiments}

This section focuses on providing empirical evaluation results trying to address the following four research questions where RQ1 presents general comparison results of {\ourmethod} with the baseline, RQ2 conducts ablation studies on different modules of {\ourmethod}, RQ3 and RQ4 focus on exploring important factors in internal hidden states and external confidence guidance, respectively: 

\begin{itemize}[noitemsep, topsep=0pt, parsep=3pt, partopsep=0pt, leftmargin=*]
    \item \textit{RQ1: How does {\ourmethod} perform across different tasks and settings?} We begin with a comprehensive comparison of our method’s performance against baseline approaches to provide a general overview.
    \item \textit{RQ2: How effective is each module of {\ourmethod} ?} Building on the results of RQ1, we conduct systematic ablation studies to evaluate the contribution of each module in our approach.
    \item \textit{RQ3: How does the choice of layer selection impact final performance?} We present empirical results to examine the impact of layer selection, which is closely tied to our method’s use of internal-state guidance.
    \item \textit{RQ4: What is the effect of our confidence aggregation method?} Finally, we analyze the performance influence of our proposed confidence-driven distribution-aware aggregation methods used in the intra-cluster sampling stage.
\end{itemize}

\subsection{RQ1: How does {\ourmethod} perform across different tasks and settings?}
\label{sec:experiments:rq1}

{The results for Llama-2 and Code Llama are presented in Table~\ref{tab:main_result_llama2}, and the results for Llama-3, Qwen2.5, and Phi-4 are shown in Table~\ref{tab:main_result_llama3}-~\ref{tab:main_result_phi}, respectively.} 
{In summary, {\ourmethod} archives SOTA performance in 26 of 32 settings.} 
This is particularly challenging, given that we are comparing different tasks with sizes spanning three orders of magnitude across five models. In particular, {\ourmethod} outperforms all baselines on Llama-3, achieving the highest estimation accuracy improvement of 37.18\% on NQ-open. {We can observe similar trends across other models. For instance, Llama-2 achieves state-of-the-art performance in 5 out of 6 cases, Code Llama in 2 out of 2 cases, Qwen2.5 in 6 out of 8 cases, and Phi-4 in 5 out of 8 cases.}


On the other hand, some baseline methods that perform well for classification models may exhibit poor performance when applied to LLMs. For instance, {\pacesampling} achieves an AUC of 0.2123 on HumanEval for Code Llama, while {\ourmethod} achieves a significantly lower AUC of 0.0300. {Another example is CESampling on AGIEval for the Phi-4 model, which achieves an AUC of 0.0203, while {\ourmethod} achieves an AUC of 0.0085.} These substantial degradations likely stem from fundamental differences between LLMs and classification DNNs. Unlike classification models, LLMs have more complex internal structures and rely on an autoregressive generation process that produces thousands of token-level confidence scores, introducing additional challenges for effective evaluation and sampling. As such, we can also conclude that the prior experience, findings, and solutions for traditional DNNs may no longer be directly applicable to LLMs and will require further adaptation.

\begin{table*}[htbp]
    \centering
    \renewcommand{\arraystretch}{1.0}
    \footnotesize
     \resizebox{0.95\columnwidth}{!}{
    \setlength{\tabcolsep}{1pt}
    \begin{tabular}{lcccccccc}
    \toprule
     & AGIEval & TriviaQA  & NQ-open & GSM8K & TruthfulQA-I &  TruthfulQA-T & HumanEval & MBPP \\
    \midrule
    \pacesampling  & 0.0275 & 0.0410 & 0.0650 & 0.0316 & 0.0067 & \toponetext{0.0111} & 0.2123 & \toptwotext{0.0305} \\
    \csssampling & 0.0200 & 0.0032 & \toptwotext{0.0201} & 0.0438 & \toptwotext{0.0037} & {0.0182} & 0.0560 & 0.0320 \\
    \cessampling & \toptwotext{0.0152} & \toptwotext{0.0031} & 0.0303 & 0.0593 & 0.0064 & 0.0253 & 0.0639 & 0.0473 \\
    
    \rhcsampling & 0.0226 & 0.0035 & 0.0202 & 0.0386 & 0.0043 & \toponetext{0.0182} & 0.0629 & {0.0305} \\
    \diffusesampling & 0.0532 & 0.0331 & 0.0544 & \toptwotext{0.0279} & 0.0070 & 0.0337 & \toptwotext{0.0418} & {0.0329}  \\
    \midrule
    \midrule
    \ourmethod & \toponetext{0.0101} & \toponetext{0.0027} & \toponetext{0.0180} & \toponetext{0.0262} & \toponetext{0.0036} & {0.0232} & \toponetext{0.0300} & \toponetext{0.0256}  \\
    \bottomrule
    \end{tabular} 
    }
    \caption{
    AUC for relative estimation error across sampling proportions from 5\% to 50\% for Llama-2 and CodeLlama. The best performance is indicated by the \toponetext{\textit{top-1}} color, and the second best by the \toptwotext{\textit{top-2}} color. 
    }
    \label{tab:main_result_llama2}
    \vspace{-5pt}
\end{table*}

\begin{table*}[htbp]
    \centering
    \footnotesize
     \resizebox{0.95\columnwidth}{!}{
    \setlength{\tabcolsep}{1pt}
    \begin{tabular}{lcccccccc}
    \toprule
     & AGIEval & TriviaQA  & NQ-open & GSM8K & TruthfulQA-I &  TruthfulQA-T & HumanEval & MBPP \\
    \midrule
    \pacesampling  & 0.0290 & 0.0091 & 0.0417 & 0.0270 & 0.0045 & 0.0262 & 0.3358 & 0.0412 \\
    \csssampling & \toptwotext{0.0151} & \toptwotext{0.0029} & \toptwotext{0.0252} & 0.0185 & 0.0037 & \toptwotext{0.0161} & {0.0804} & 0.0393 \\
    \cessampling & 0.0163 & 0.0030 & 0.0269 & 0.0241 & \toptwotext{0.0035} & 0.0194 & 0.0921 & \toptwotext{0.0377} \\
    
    \rhcsampling & 0.0236 & 0.0032 & 0.0283 & 0.0918 & 0.0112 & 0.0217 & 0.1408 & 0.0547  \\
    \diffusesampling & 0.0396 & 0.0272 & 0.0596 & \toptwotext{0.0184} & 0.0178 & 0.0834 & \toptwotext{0.0772} & 0.0415  \\
    \midrule
    \midrule
    \ourmethod & \toponetext{0.0104} & \toponetext{0.0024} & \toponetext{0.0158} & \toponetext{0.0164} & \toponetext{0.0034} & \toponetext{0.0135} & \toponetext{0.0659} & \toponetext{0.0320}  \\
    \bottomrule
    \end{tabular}
    }
    \caption{
    AUC for relative estimation error across sampling proportions from 5\% to 50\% for Llama-3. We additionally report our method's performance gain compared with the best method in baselines. The best performance is indicated by the \toponetext{\textit{top-1}} color, and the second best by the \toptwotext{\textit{top-2}} color. 
    }
    \vspace{-5pt}
    \label{tab:main_result_llama3}
\end{table*}

\begin{table*}[htbp]
    \centering
    \footnotesize
    
    \resizebox{0.95\columnwidth}{!}{
    \setlength{\tabcolsep}{1pt}
    \begin{tabular}{lcccccccc}
    \toprule
                      & AGIEval & TriviaQA  & NQ-open & GSM8K & TruthfulQA-I &  TruthfulQA-T & HumanEval & MBPP  \\
\midrule
\pacesampling      & 0.0223 & 0.0253 & 0.0279 & 0.0131 & \toptwotext{0.0048} & 0.0189 & 0.1554 & 0.0199  \\
\csssampling       & \toptwotext{0.0095} & 0.0040 & 0.0200 & 0.0076 & 0.0051 & 0.0101 & \toponetext{0.0411} & {0.0191}  \\
\cessampling    & 0.0104 & \toptwotext{0.0031} & \toptwotext{0.0184} & 0.0180 & \toponetext{0.0041} & \toptwotext{0.0092} & \toptwotext{0.0412} & 0.0195 \\
\rhcsampling       & 0.0158 & 0.0048 & 0.0232 & \toptwotext{0.0069} & 0.0084 & 0.0096 & 0.0441 & \toptwotext{0.0184} \\
\diffusesampling   & 0.0498 & 0.0308 & 0.0380 & 0.0103 & 0.0072 & 0.0160 & 0.0533 & 0.0345 \\
\midrule
\midrule
\ourmethod & \toponetext{0.0081} & \toponetext{0.0029} & \toponetext{0.0178} & \toponetext{0.0058} & 0.0053 & \toponetext{0.0039} & 0.0565 & \toponetext{0.0130} \\
\bottomrule
    \end{tabular}
    }
        \caption{ {  AUC for relative estimation error across sampling proportions from 5\% to 50\% for Qwen2.5. We additionally report our method's performance gain compared with the best method in baselines. The best performance is indicated by the \toponetext{\textit{top-1}} color, and the second best by the \toptwotext{\textit{top-2}} color. 
        }
        }

    \label{tab:main_result_qwen}
\end{table*}
\begin{table*}[htbp]
    \centering
    \footnotesize
    
    \resizebox{0.95\columnwidth}{!}{
    \setlength{\tabcolsep}{1pt}
    \begin{tabular}{lcccccccc}
    \toprule
                      & AGIEval & TriviaQA  & NQ-open & GSM8K & TruthfulQA-I &  TruthfulQA-T & HumanEval & MBPP  \\
    \midrule
\pacesampling      & 0.0305                      & 0.0595                       & 0.0340                        & 0.0177                    & \toponetext{0.0035}                                      & 0.0065                                       & 0.1391                        & 0.0256                   \\
\csssampling       & \toptwotext{0.0088}                      & {0.0040}                        & 0.0152                       & 0.0071                    & \toptwotext{0.0038}                                      & \toptwotext{0.0063}                                       & 0.0443                        & 0.0178                   \\
\cessampling    & 0.0203                      & \toptwotext{0.0035}                       & 0.0167                       & 0.0138                    & 0.0045                                      & 0.0080                                        & 0.0640                         & \toptwotext{0.0175}                   \\
\rhcsampling       & 0.0096                      & 0.0053                       & 0.0178                       & \toptwotext{0.0067}                    & 0.0042                                      & 0.0071                                       & 0.0470                         & 0.0178                   \\
\diffusesampling   & 0.0359                      & 0.0460                        & \toponetext{0.0067}                       & 0.0096                    & 0.0039                                      & 0.0064                                       & \toponetext{0.0326}                        & 0.0221                   \\
\midrule
\midrule 
\ourmethod & \toponetext{0.0085}                      & \toponetext{0.0032}                       & \toptwotext{0.0128}                       & \toponetext{0.0053}                    & 0.0040                                       & \toponetext{0.0059}                                       & \toptwotext{0.0411}                        & \toponetext{0.0164} \\                    
\bottomrule
    \end{tabular}
    }
        \caption{
        { AUC for relative estimation error across sampling proportions from 5\% to 50\% for Phi-4. We additionally report our method's performance gain compared with the best method in baselines. The best performance is indicated by the \toponetext{\textit{top-1}} color, and the second best by the \toptwotext{\textit{top-2}} color.} 
        }

    \label{tab:main_result_phi}
\end{table*}

However, we can also observe some exception cases for {\ourmethod}.  
{A representative type is its performance on TruthfulQA-I and TruthfulQA-T. It exhibits suboptimal performance on TruthfulQA-T with Llama-2 and on TruthfulQA-I with Qwen2.5 and Phi-4. Upon further investigation, we hypothesize that this anomaly might arise from the fact that truthfulness and performance are fundamentally different properties. They may exhibit entirely different internal and external behavior patterns and are likely to be model-specific.}

{Previous studies have indicated that internal states can partially reflect a model's truthfulness~\cite{azaria2023the} or knowledge~\cite{ni2025towards}. However, capturing this information explicitly across different LLMs may require specialized techniques, such as training classifiers or using prompts to activate truthfulness- or knowledge-related neurons~\cite{zou2023representation}. In models like Phi-4 and Qwen2.5, neurons related to truthfulness might be more active, while they are less so in Llama-2. As a result, the PCA used in {\ourmethod} captures different information for each model. Since our approach did not incorporate these specialized methods, the internal states may not have effectively guided the sampling process in this context.}


{Another possible explanation could be the misalignment between the models' confidence and the informativeness or truthfulness of a given model. In Llama-2, confidence might be more closely correlated with informativeness, whereas in Qwen2.5 and Phi-4, it might be more aligned with truthfulness.}
As such, our output-driven intra-cluster search is also misled. 

Overall, it is worth mentioning that {\ourmethod} is a general framework as its performance can be further improved by incorporating more advanced neuron extraction and analysis techniques, more intelligent inter-cluster dynamic sampling strategies, and more precise intra-cluster optimization-based test point selection methods.

\begin{finding}
    \label{finding:1}
    {\ourmethod} achieves state-of-the-art performance in {26 out of 32} cases across {five} models and eight settings. After more in-depth investigations into the experiment results, we found that approaches and insights derived from classical DNNs may no longer be effective for LLMs. Additionally, our findings also highlight the necessity of developing more adaptive methods based on {\ourmethod} to accommodate the diverse behavioral patterns of {different} LLMs across various tasks.
\end{finding}

\subsection{RQ2: How effective is each module of {\ourmethod}?}
\label{sec:experiments:rq2}

In this section, we conduct ablation studies to assess and understand the effectiveness of each component in {\ourmethod}'s framework design.
In particular, we focus on three components: 
(1) The automatic cluster number search algorithm in charge of extracting the underlying LLMs' behavior patterns in the test space (CluSearch); 
(2) The adaptive strategy that utilizes stratified sampling based on the computation of Monte Carlo Upper Confidence Bound (MC-UCB); 
(3) The sub-sampling strategy inside each cluster to preserve an unbiased data selection with respect to LLMs' confidence distribution (SubSample).  
Corresponding to the three key components of {\ourmethod}, we establish three distinct configurations for our ablation studies:

\begin{itemize}[noitemsep, topsep=0pt, parsep=3pt, partopsep=0pt, leftmargin=*]
    \item {\ourmethod} without CluSearch ({\ourmethod}-CluSearch).
    We employ a fixed number of clusters\footnote{We set the cluster number as 8 according to the sklearn K-means default parameter.} for the unsupervised learning algorithm across all datasets and sampling proportions.
    \item {\ourmethod} without adaptive stratified sampling ({\ourmethod}-Inter-cluster Search)
    In this setting, we replace the adaptive approach with random sampling 
    \item {\ourmethod} without the sub-sampling strategy within each cluster ({\ourmethod}-Intra-cluster Search).
    When a cluster is selected by the MC-UCB, we randomly choose a point within the cluster for sampling.
\end{itemize}

\begin{table}[h]
    \centering
    \renewcommand{\arraystretch}{1.0}
     \resizebox{0.85\columnwidth}{!}{
    \setlength{\tabcolsep}{2pt}
    \begin{tabular}{lcccccccc}
    \toprule
     & AGIEval & TriviaQA  & NQ-open & GSM8K & TQA-I &  TQA-T & HumanEval & MBPP \\
    \midrule
    \multirow{2}{*}{-CluSearch} & 0.0227 & 0.00285 & 0.0212 & 0.0384 & 0.0056 & 0.0252 & 0.1021 & 0.0298  \\
     & -124.51\% & -6.53\% & -17.64\% & -46.74\% & -54.24\% & -8.44\% & -239.65\% & -16.19\%  \\

     \midrule

    \multirow{2}{*}{-Inter} & 0.0135 & 0.00294 & 0.017 & 0.025 & 0.0037 & 0.0205 & 0.0467 & 0.0274 \\
    
    & -34.08\% & -9.97\% & +5.27\% & +4.58\% & -3.28\% & +11.80\% & -55.45\% & -6.75\% \\
    
    \midrule
    
    \multirow{2}{*}{-Intra} & 0.0168 & 0.0027 & 0.0156 & 0.0351 & 0.0036 & 0.0195 & 0.0621 & 0.0325 \\
    
    & -66.58\% & -0.67\% & -13.54\% & -34.21\% & + 0.53\% & + 15.96\% & -106.72\% & -26.98\%  \\
    \bottomrule
    \end{tabular}
     }
    \caption{
        Ablation study results on Llama-2 and Code Llama with relative performance difference.
    }
    \label{tab:ablation_study_llama2}
\end{table}

\begin{table}[h]
    \centering
    \renewcommand{\arraystretch}{1.0}
    \small
     \resizebox{0.85\columnwidth}{!}{
    \setlength{\tabcolsep}{2pt}
    \begin{tabular}{lcccccccc}
    \toprule
     & AGIEval & TriviaQA  & NQ-open & GSM8K & TQA-I &  TQA-T & HumanEval & MBPP \\
    \midrule
    \multirow{2}{*}{-CluSearch} & 0.0108 & 0.0030 & 0.0246 & 0.0186 & 0.0028 & 0.0171 & 0.1552 & 0.0388  \\
     & -3.46\% & -25.42\% & -56.00\% & -13.21\% & +16.53\% & -26.92\% & -135.51\% & -21.36\%  \\

     \midrule

    \multirow{2}{*}{-Inter} & 0.0135 & 0.0025 & 0.0141 & 0.0193 & 0.0038 & 0.0185 & 0.0676 & 0.0373 \\
    
    & -30.26\% & -3.08\% & +10.75\% & -17.98\% & -11.26\% & -36.87\% & -2.63\% & -16.46\% \\
    
    \midrule
    
    \multirow{2}{*}{-Intra} & 0.0118 & 0.0029 & 0.0170 & 0.0250 & 0.0037 & 0.0205 & 0.0467 & 0.0274 \\
    
    & -13.51\% & -22.75\% & -7.84\% & -52.18\% & -9.32\% & -51.78\% & +29.12\% & +14.52\%  \\
    \bottomrule
    \end{tabular}
     }
    \caption{
        Ablation study results on Llama-3 with relative performance difference.
    }
    \label{tab:ablation_study_llama3}
\end{table}

\begin{table}[h]
    \centering
    \renewcommand{\arraystretch}{1.0}
    
    \small
     \resizebox{0.85\columnwidth}{!}{
    \setlength{\tabcolsep}{2pt}
    \begin{tabular}{lcccccccc}
    \toprule
     & AGIEval & TriviaQA  & NQ-open & GSM8K & TQA-I &  TQA-T & HumanEval & MBPP \\
    \midrule
    \multirow{2}{*}{-CluSearch} & 0.0130 & 0.0039 & 0.0200 & 0.0059 & 0.0043 & 0.0069 & 0.1088 & 0.0198  \\
     & -60.85\% & -35.77\% & -12.73\% & -3.04\% & +18.80\% & -77.52\% & -92.60\% & -52.37\%
  \\

     \midrule

    \multirow{2}{*}{-Inter} & 0.0096 & 0.0030 & 0.0198 & 0.0045 & 0.0054 & 0.0152 & 0.0803 & 0.0204 \\
    
    & -19.06\% & -4.17\% & -11.41\% & +22.63\% & -1.95\% & -293.39\% & -42.20\% & -56.52\%
 \\
    
    \midrule
    
    \multirow{2}{*}{-Intra} & 0.0084 & 0.0034 & 0.0178 & 0.0075 & 0.0048 & 0.0199 & 0.0449 & 0.0194 \\
    
    & -3.95\% & -16.16\% & -0.28\% & -29.26\% & +10.14\% & -413.45\% & +20.49\% & -49.04\%
  \\
    \bottomrule
    \end{tabular}
     }
    \caption{
        {Ablation study results on Qwen2.5 with relative performance difference.}
    }
    \label{tab:ablation_study_qwen}
\end{table}

\begin{table}[h]
    \centering
    \renewcommand{\arraystretch}{1.0}
    
    \small
     \resizebox{0.85\columnwidth}{!}{
    \setlength{\tabcolsep}{2pt}
    \begin{tabular}{lcccccccc}
    \toprule
     & AGIEval & TriviaQA  & NQ-open & GSM8K & TQA-I &  TQA-T & HumanEval & MBPP \\
    \midrule
    \multirow{2}{*}{-CluSearch} & 0.0102 & 0.0049 & 0.0137 & 0.0054 & 0.0037 & 0.0127 & 0.1146 & 0.0227  \\
     & -19.86\% & -52.94\% & -6.73\% & -1.98\% & +8.68\% & -115.03\% & -178.94\% & -38.71\%
  \\

     \midrule

    \multirow{2}{*}{-Inter} & 0.0097 & 0.0037 & 0.0168 & 0.0073 & 0.0047 & 0.0057 & 0.0457 & 0.0179 \\
    
    & -14.60\% & -15.69\% & -31.30\% & -36.85\% & -16.60\% & +3.63\% & -11.22\% & -9.10\% \\
    
    \midrule
    
    \multirow{2}{*}{-Intra} & 0.0090 & 0.0056 & 0.0117 & 0.0045 & 0.0045 & 0.0041 & 0.0735 & 0.0174 \\
    
    & -6.38\% & -75.09\% & +8.38\% & +15.98\% & -11.65\% & +30.05\% & -78.83\% & -6.19\%
  \\
    \bottomrule
    \end{tabular}
     }
    \caption{
        {Ablation study results on Phi-4 with relative performance difference.}
    }
    \label{tab:ablation_study_phi}
\end{table}

{We conducted experiments on five models across 24 different settings, keeping all other configurations consistent with the previous experiments. The corresponding results for Llama-2 and Code Llama are presented in Table~\ref{tab:ablation_study_llama2}, the results for Llama-3, Qwen2.5 and Phi-4 are shown in Table~\ref{tab:ablation_study_llama3}-~\ref{tab:ablation_study_phi}.}
We noticed that all three components are essential for {\ourmethod} in most settings. 
The most significant influence is CluSearch. {Removing CluSearch will result in an average 43.02\% performance drop on Llama-2, 127.92\% on Code Llama, 33.17\% on Llama-3, 39.51\% on Qwen2.5, and 50.69\% on Phi-4. Among all the cases in the ClusSearch ablation study, Code Llama on HumanEval is the most significant, resulting in a 239.65\% drop in estimation accuracy, making the method much less effective. }

On the other hand, our intra-cluster sampling method also significantly contributes to the final performance in certain scenarios. {For instance, without this technique, we observed a performance drop of up to 413.45\% on Qwen2.5 for TQA-T and 106.72\% on Code Llama for HumanEval.} 
These results highlight the importance of the intra-cluster sampling method. 
{Finally, our inter-cluster sampling strategy contributes modestly across all five models. With the exception of Qwen2.5 on TQA-T, which experiences a 293.39\% drop, removing this strategy will result in a performance decrease of no more than 56\%. Despite its moderate impact, this strategy enhances performance in 26 out of 32 cases, demonstrating its broad applicability and consistent benefits.} While its impact is not as pronounced as the other two components, this strategy still plays a vital role in terms of the overall performance of {\ourmethod}.


\begin{finding}
    \label{finding:2}
    All three components are essential for accurate test estimation capability of {\ourmethod}. Among them, CluSearch has the most significant impact on performance, followed by Intra-cluster sampling, which provides notable contributions. Inter-cluster sampling, while less impactful, still delivers moderate improvements.
\end{finding}


\subsection{RQ3: How does the choice of layer selection impact final performance?}
\label{sec:experiments:rq3}

A crucial design choice in most internal-state-driven approaches is selecting which layer to use as the feature source. Much like the human brain, different layers within LLMs are responsible for distinct functionalities~\cite{jin2024exploring}. Consequently, this choice can have a significant impact on the model’s final performance.

Previous studies on classification models typically favor selecting the last hidden layer, assuming it has the most pertinent information regarding their decisions~\cite{li2019boosting, bengio2013better, feinman2017detecting, kim2019guiding, wang2022exploratory}. However, recent research on LLMs indicates that the optimal layer for feature extraction varies depending on the task, with intermediate layers often demonstrating superior performance in various downstream application~\cite{azaria2023the, zou2023representation, gurnee2024language, chen2024inside}. 

\begin{table*}[htbp]
    \centering
    \renewcommand{\arraystretch}{1.0}
     \resizebox{0.9\columnwidth}{!}{
    \setlength{\tabcolsep}{2pt}
    \begin{tabular}{lcccccccc}
    \toprule
     & AGIEval & TriviaQA  & NQ-open & GSM8K & TruthfulQA-I &  TruthfulQA-T & HumanEval & MBPP \\
    \midrule
    Middle Layer & 0.0101 & 0.0027 & 0.0180 & 0.0262 & 0.0036 & 0.0232 & 0.0300 & 0.0256 \\ \midrule
    
    \multirow{2}{*}{Last Layer} & 0.0090 & 0.0030 & 0.0339 & 0.0324 & 0.0054 & 0.0209 & 0.0436 & 0.0208 \\
        & +10.75\% & -12.88\% & -88.29\% & -23.83\% & -50.15\% & +10.00\% & -45.12\% & +18.67\% \\
        
    
    \hline
    \end{tabular}
     }
    \caption{
        Target layer selection with performance on different datasets and relative performance variation on Llama-2 and Code Llama. 
    }\label{tab:layer_ablation_llama2}
\end{table*}

\begin{table*}[htbp]
    \centering
    \renewcommand{\arraystretch}{1.0}
     \resizebox{0.9\columnwidth}{!}{
    \setlength{\tabcolsep}{2pt}
    \begin{tabular}{lcccccccc}
    \toprule
     & AGIEval & TriviaQA  & NQ-open & GSM8K & TruthfulQA-I &  TruthfulQA-T & HumanEval & MBPP \\
    \midrule
    Middle Layer & 0.0108 & 0.0024 & 0.0158 & 0.0164 & 0.0034 & 0.0135 & 0.0659 & 0.032 \\ \midrule
    
    \multirow{2}{*}{Last Layer} & 0.0139 & 0.0026 & 0.0204 & 0.0187 & 0.0031 & 0.0166 & 0.0673 & 0.0426 \\
        & -33.92\% & -7.60\% & -28.87\% & -13.88\% & +9.43\% & -22.81\% & -2.12\% & -33.30\% \\
        
    
    \hline
    \end{tabular}
     }
    \caption{
        Target layer selection with performance on different datasets and relative performance variation on Llama-3. 
    }\label{tab:appendix_layer_ablation_llama3}
\end{table*}

\begin{table*}[htbp]
    \centering
    \renewcommand{\arraystretch}{1.0}
    
     \resizebox{0.9\columnwidth}{!}{
    \setlength{\tabcolsep}{2pt}
    \begin{tabular}{lcccccccc}
    \toprule
     & AGIEval & TriviaQA  & NQ-open & GSM8K & TruthfulQA-I &  TruthfulQA-T & HumanEval & MBPP \\
    \midrule

    Middle Layer & 0.0081 & 0.0029 & 0.0178 & 0.0058 & 0.0053 & 0.0039 & 0.0565 & 0.0130 \\ \midrule
    
    \multirow{2}{*}{Last Layer} & 0.0103 & 0.0035 & 0.0211 & 0.0054 & 0.0037 & 0.0078 & 0.1174 & 0.0152 \\
        & -27.36\% & -22.16\% & -18.82\% & +6.40\% & +29.82\% & -100.34\% & -107.88\% & -16.97\% \\
    \hline
    \end{tabular}
     }
    \caption{
        {Target layer selection with performance on different datasets and relative performance variation on Qwen2.5.}
    }\label{tab:appendix_layer_ablation_qwen}
\end{table*}

\begin{table*}[htbp]
    \centering
    \renewcommand{\arraystretch}{1.0}
    
     \resizebox{0.85\columnwidth}{!}{
    \setlength{\tabcolsep}{2pt}
    \begin{tabular}{lcccccccc}
    \toprule
     & AGIEval & TriviaQA  & NQ-open & GSM8K & TruthfulQA-I &  TruthfulQA-T & HumanEval & MBPP \\
    \midrule

    Middle Layer & 0.0085 & 0.0032 & 0.0128 & 0.0053 & 0.0040 & 0.0059 & 0.0411 & 0.0164
 \\ \midrule
    
    \multirow{2}{*}{Last Layer} & 0.0090 & 0.0056 & 0.0117 & 0.0045 & 0.0045 & 0.0041 & 0.0735 & 0.0174 \\
        & -6.38\% & -75.09\% & +8.38\% & +15.98\% & -11.65\% & +30.05\% & -78.83\% & -6.19\%
 \\
    \hline
    \end{tabular}
     }
    \caption{
        {Target layer selection with performance on different datasets and relative performance variation on Phi-4.}
    }\label{tab:appendix_layer_ablation_phi}
\end{table*}

To identify the most effective layer, we performed ablation studies on {all the five LLMs}. The study involves repeating all the experiments from Section~\ref{sec:experiments:rq1} while varying only the target layer, allowing us to assess the impact of layer selection on performance. {We compare the results on the middle layer and the last layer in Table~\ref{tab:layer_ablation_llama2} to Table~\ref{tab:appendix_layer_ablation_phi}}.

The result shows that the middle layer is the most straightforward layer to choose as the target feature layer. {As two representative examples, switching from the middle layer to the last layer resulted in significant performance drops on HumanEval on Qwen2.5 and NQ-open for Llama-2, with decreases of 107.88\% and 88.29\%, respectively.} The results indicate that selecting the middle layer generally leads to superior performance. 
{However, there are still some cases in which the middle layer is better. one case each for Llama-2, Code Llama, and Llama-3; two cases for Qwen2.5; and three cases for Phi-4. Notably, 6 out of these 8 instances occur on the GSM8K and TruthfulQA datasets.}
This suggests potential for improvement by developing algorithms that dynamically select feature layers tailored to specific tasks, which could enhance performance further in future work. 

\begin{finding}
    \label{finding:3}
    Contrary to the common belief that the final layer plays the most critical role in a model’s decision-making process, our findings suggest that the middle layer is more effective in guiding the sampling process and achieving accurate performance estimation. {Switching from the middle layer to the final layer leads to an average accuracy drop of 21.72\% across the five LLMs.}
\end{finding}


\subsection{RQ4: What is the effect of our confidence aggregation method?}
\label{sec:experiments:rq4}

Different from pre-LLM DNNs, which usually produce a single confidence score, LLMs generate outputs consisting of up to thousands of tokens. Aggregating information from these tokens into a single confidence score is a crucial design choice. This decision can significantly impact the effectiveness of output-guided sampling. 

In this study, we propose entropy-guided selection criteria for four different kinds of aggregation methods, namely, selecting the first token (first), computing the maximum confidence over the entire generation sequence(\textit{max}), computing the average confidence over the entire sequence(\textit{mean}), computing the geometric mean over the entire sequence (\textit{gmean}). We divide the confidence interval into a histogram consisting of 10 bins and calculate the entropy. {The corresponding results for Llama-2 and Code Llama are shown in Table~\ref{tab:llm_entropy_llama2}, and results for Llama-3, Qwen2.5, and Phi-4 are shown in Table~\ref{tab:llm_entropy_llama3}-~\ref{tab:llm_entropy_phi}.} It is clear that the confidence score distribution varies significantly across different datasets and different models with different aggregation methods. For example, when evaluating Code Llama on the HumanEval dataset, the entropy of the \textit{max} aggregation is zero (i.e., falling into the same bin), while it is 2.733 for the \textit{first} aggregation. In {\ourmethod}, We select the aggregation method according to the maximum entropy across four different methods for each dataset and each model.

\begin{table*}[h] 
    \centering
    \renewcommand{\arraystretch}{1.0}
    \resizebox{0.9\columnwidth}{!}{
    \begin{tabular}{cccccccccc}
    \toprule
         Dataset & AGIEval & TrivialQA & NQ-OPEN & GSM8K & TruthfulQA-I & TruthfulQA-T & HumanEval & MBPP  \\ \midrule
         
        
        first & 2.121 & 3.298 & 2.269 & 0.683 & -0.000 & -0.000 & 2.733 & 0.193 \\ 

        gmean & 2.725 & 2.996 & 2.941 & 0.448 & 2.072 & 2.072 & 2.164 & 1.738 \\
        
        max & 0.645 & 1.456 & 1.492 & 0.034 & -0.000 & -0.000 & -0.000 & -0.000 \\

        mean & 2.634 & 2.801 & 2.847 & 0.398 & 1.757 & 1.757 & 1.763 & 1.234 \\
        \bottomrule
    \end{tabular}
    }
    \caption{Entropy of different aggregation methods for Llama-2 and Code Llama }
    \label{tab:llm_entropy_llama2}
\end{table*}

\begin{table*}[h] 
    \centering
    \renewcommand{\arraystretch}{1.0}
    \resizebox{0.9\columnwidth}{!}{
    \begin{tabular}{cccccccccc}
    \toprule
         Dataset & AGIEval & TrivialQA & NQ-OPEN & GSM8K & TruthfulQA-I & TruthfulQA-T & HumanEval & MBPP  \\ \midrule
         
        
        first & 1.219 & 0.409 & 1.309 & 1.341 & -0.000 & -0.000 & 2.554 & 2.469 \\ 

        gmean & 2.550 & 0.920 & 2.300 & 0.087 & 2.470 & 2.470 & 2.656 & 2.311 \\
        
        max & 2.386 & 2.281 & 1.673 & 1.604 & -0.000 & -0.000 & 0.676 & 0.021 \\

        mean & 3.096 & 2.805 & 2.845 & 0.688 & 2.212 & 2.212 & 2.542 & 1.797 \\
        \bottomrule
    \end{tabular}
    }
    \caption{Entropy of different aggregation methods for Llama-3 }
    \label{tab:llm_entropy_llama3}
\end{table*}

\begin{table*}[h] 
    \centering
    
    \renewcommand{\arraystretch}{1.0}
    \resizebox{0.9\columnwidth}{!}{
    \begin{tabular}{cccccccccc}
    \toprule
         Dataset & AGIEval & TrivialQA & NQ-OPEN & GSM8K & TruthfulQA-I & TruthfulQA-T & HumanEval & MBPP  \\ \midrule
                 
        first & 1.270 & 0.404 & 1.528 & 0.080 & 0.014 & 0.014 & 2.203 & 1.820 \\ 

        gmean & 3.023 & 1.114 & 1.585 & 0.215 & 0.903 & 0.903 & 2.760 & 1.607 \\
        
        max & 0.926 & 2.288 & 2.009 & 0.000 & 0.587 & 0.587 & 0.255 & 0.000 \\

        mean & 2.744 & 2.944 & 2.903 & 0.170 & 0.854 & 0.854 & 2.455 & 1.362 \\
        \bottomrule
    \end{tabular}
    }
    \caption{{Entropy of different aggregation methods for Qwen2.5 }}
    \label{tab:llm_entropy_qwen}
\end{table*}

\begin{table*}[h] 
    \centering
    
    \renewcommand{\arraystretch}{1.0}
    \resizebox{0.9\columnwidth}{!}{
    \begin{tabular}{cccccccccc}
    \toprule
         Dataset & AGIEval & TrivialQA & NQ-OPEN & GSM8K & TruthfulQA-I & TruthfulQA-T & HumanEval & MBPP  \\ \midrule
                 
        first & 1.948 & 2.921 & 1.657 & 0.650 & 2.666 & 2.666 & 1.625 & 1.873 \\ 

        gmean & 2.914 & 2.498 & 1.788 & 0.638 & 2.207 & 2.207 & 2.433 & 1.458 \\
        
        max & 0.631 & 0.994 & 2.125 & 0.009 & 0.000 & 0.000 & 0.219 & 0.000 \\

        mean & 2.488 & 2.325 & 2.894 & 0.511 & 1.658 & 1.658 & 1.870 & 1.185 \\
        \bottomrule
    \end{tabular}
    }
    \caption{{Entropy of different aggregation methods for Phi-4 }}
    \label{tab:llm_entropy_phi}
\end{table*}

\begin{table*}[h] 
    \centering
    \renewcommand{\arraystretch}{1.0}
    \resizebox{0.9\columnwidth}{!}{
    \begin{tabular}{cccccccccc}
    \toprule
         Dataset & AGIEval & TrivialQA & NQ-OPEN & GSM8K & TruthfulQA-I & TruthfulQA-T & HumanEval & MBPP  \\ \midrule
         
        
        first & 0.920 & 0 & 0.969 & 0.725 & 0.777 & 0.777 & 0.086 & 0.975 \\ 

        gmean & 0.511 & 0.998 & 0.903 & 0.998 & 0.991 & 0.991 & 0.992 & 0.991 \\
        
        max & 0.000 & 1.000 & 0.037 & 0.029 & 0.757 & 0.757 & 0.992 & 0.994 \\

        mean & 0.000 & 0.944 & 0.585 & 0.999 & 0.955 & 0.955 & 0.946 & 0.988 \\
        \bottomrule
    \end{tabular}
    }
    \caption{Dip Test P-value for different aggregation methods on Llama-2 and Code-Llama}
    \label{tab:llm_diptest_llama2}
\end{table*}

\begin{table*}[h] 
    \centering
    \renewcommand{\arraystretch}{1.0}
    \resizebox{0.9\columnwidth}{!}{
    \begin{tabular}{cccccccccc}
    \toprule
         Dataset & AGIEval & TrivialQA & NQ-OPEN & GSM8K & TruthfulQA-I & TruthfulQA-T & HumanEval & MBPP  \\ \midrule
         
        
        first & 0.000 & 0.065 & 0.999 & 0.914 & 0.693 & 0.693 & 0.830 & 0.000 \\ 

        gmean & 0.000 & 0.992 & 0.652 & 0.994 & 0.001 & 0.001 & 0.025 & 0.992 \\
        
        max & 0.171 & 0.000 & 0.022 & 0.993 & 0.954 & 0.954 & 0.831 & 0.993 \\

        mean & 0.000 & 0.000 & 0.199 & 0.992 & 0.009 & 0.009 & 0.720 & 0.946 \\
        \bottomrule
    \end{tabular}
    }
    \caption{Dip Test P-value for different aggregation methods on Llama-3}
    \label{tab:llm_diptest_llama3}
\end{table*}

\begin{table*}[h] 
    \centering
    
    \renewcommand{\arraystretch}{1.0}
    \resizebox{0.9\columnwidth}{!}{
    \begin{tabular}{cccccccccc}
    \toprule
         Dataset & AGIEval & TrivialQA & NQ-OPEN & GSM8K & TruthfulQA-I & TruthfulQA-T & HumanEval & MBPP  \\ \midrule
        
        first & 0.998 & 0.001 & 0.000 & 0.995 & 0.992 & 0.992 & 0.963 & 0.995 \\ 

        gmean & 0.000 & 0.996 & 1.000 & 0.992 & 0.058 & 0.058 & 0.953 & 0.953 \\
        
        max & 0.991 & 0.000 & 0.000 & 0.871 & 0.000 & 0.000 & 0.993 & 0.991 \\

        mean & 0.996 & 0.000 & 0.000 & 0.992 & 0.000 & 0.000 & 0.962 & 0.682 \\
        \bottomrule
    \end{tabular}
    }
    \caption{{Dip Test P-value for different aggregation methods on Qwen2.5}}
    \label{tab:llm_diptest_qwen}
\end{table*}

\begin{table*}[h] 
    \centering
    
    \renewcommand{\arraystretch}{1.0}
    \resizebox{0.9\columnwidth}{!}{
    \begin{tabular}{cccccccccc}
    \toprule
         Dataset & AGIEval & TrivialQA & NQ-OPEN & GSM8K & TruthfulQA-I & TruthfulQA-T & HumanEval & MBPP  \\ \midrule
        
        first & 1.000 & 0.571 & 0.003 & 1.000 & 0.819 & 0.819 & 0.937 & 0.998 \\ 

        gmean & 0.820 & 1.000 & 0.843 & 0.994 & 0.991 & 0.991 & 0.839 & 0.898 \\
        
        max & 0.999 & 0.413 & 0.000 & 0.000 & 0.000 & 0.000 & 0.885 & 0.995 \\

        mean & 0.229 & 1.000 & 0.000 & 0.994 & 0.980 & 0.980 & 0.981 & 0.985 \\
        \bottomrule
    \end{tabular}
    }
    \caption{{Dip Test P-value for different aggregation methods on Phi-4}}
    \label{tab:llm_diptest_phi}
\end{table*}

\begin{table}[h!]
    \centering
    
    \resizebox{0.6\columnwidth}{!}{
    \begin{tabular}{l|c|c|c|c}
    \hline
    \multirow{2}{*}{\textbf{Dataset}} & \multicolumn{1}{c|}{\textbf{Llama-2}} & \multirow{2}{*}{\textbf{Llama-3}} &  \multirow{2}{*}{\textbf{Qwen2.5}} & \multirow{2}{*}{\textbf{Phi-4}} \\ 
                                      & \multicolumn{1}{c|}{\textbf{Code Llama}}  &  & &                                         \\ \hline
    {AGIEval}                 & \textit{mean} + \WD                      & \textit{first} + \WD & \textit{gmean} + \WD  & \textit{gmean} + \KS                     \\ 
    {TrivialQA}               & \textit{first} + \WD                     & \textit{mean} + \WD  & \textit{mean} + \WD & \textit{first} + \KS                    \\ 
    {NQ-open}                 & \textit{gmean} + \KS                     & \textit{mean} + \KS & \textit{mean} + \WD &  \textit{mean} + \WD                   \\ 
    {GSM8K}                   & \textit{first} + \KS                     & \textit{max} + \KS   & \textit{gmean} + \KS  & \textit{first} + \KS                       \\ 
    {TruthfulQA-I}            & \textit{gmean} + \KS                     & \textit{gmean} + \WD & \textit{gmean} + \KS & \textit{first} + \KS                     \\ 
    {TruthfulQA-T}            & \textit{gmean} + \KS                     & \textit{gmean} + \WD & \textit{gmean} + \KS & \textit{first} + \KS                     \\ 
    {HumanEval}               & \textit{first} + \KS                     & \textit{gmean} + \WD & \textit{gmean} + \KS  & \textit{gmean} + \KS                     \\ 
    {MBPP}                    & \textit{gmean} + \KS                     & \textit{first} + \WD & \textit{first} + \KS  & \textit{first} + \KS                     \\ \hline
    \end{tabular}
    }
    \caption{{Selected metrics for Llama-2, Code Llama, Llama-3, Qwen2.5 and Phi-4 on various datasets.}}
    \label{tab:selection_metrics}
\end{table}

For confidence distribution distance measurement, we use the Wasserstein distance metric and Kolmogorov-Smirnov statistics. As we discussed in Section~\ref{sec:Methodology:intra}, we perform the dip test to assess the distribution's modality and choose Wasserstein distance for multimodal distribution and Kolmogorov-Smirnov statistics for unimodal distribution. {The p-value of the dip test results of Llama-2 and Code Llama are shown in Table~\ref{tab:llm_diptest_llama2}, and the results of Llama-3, Qwen2.5 and Phi-4 are shown in Table~\ref{tab:llm_diptest_llama3}-~\ref{tab:llm_diptest_phi}.} We set the confidence as 95\%. For a p-value smaller than 5\%, we take it as a multimodal distribution and vice versa. 

{Combing with our entropy distribution results in Table~\ref{tab:llm_entropy_llama2}-\ref{tab:llm_entropy_phi} and dip test results in Table~\ref{tab:llm_diptest_llama2}-\ref{tab:llm_diptest_phi}, we conclude our final aggregation and distribution distance metric selection in Table~\ref{tab:selection_metrics}.} To highlight the significance of these design choices, we conducted two additional series of ablation studies. In the first series, we reversed the distance metric selection determined by the dip test while retaining the entropy-guided aggregation method. In the second series, we reversed the confidence aggregation method, changing it from using the highest entropy to the lowest entropy while keeping the dip test for distance metric selection unchanged. {The results for Llama-2 and Code Llama are presented in Table~\ref{tab:diptest_ablation_llama2}, and the findings for Llama-3, Qwen2.5, and Phi-4 are detailed in Table~\ref{tab:diptest_ablation_llama3}-~\ref{tab:diptest_ablation_phi}.}

The results across the five models demonstrate that both the choice of distance metric and the confidence aggregation method are critical to the overall performance of {\ourmethod} for different models and datasets. {While reversing the distance metric occasionally yields limited improvements, as observed in 11 out of 32 cases with a maximum increase of 42.93\% for Qwen2.5 on the HumanEval dataset, it significantly degrades performance in the majority of cases (21 out of 32), with a decline of up to 323.52\% for Qwen2.5 on the AGIEVAL dataset.} 
{A similar pattern is observed with the confidence aggregation methods. Although reversing the aggregation method can result in a small accuracy gain in 9 out of 32 cases, with the maximum as +25.25\% for Phi-4 on TruthfulQA-I, it causes a substantial performance drop in other 23 cases with up to -495.15\% on TruthfulQA-T. Furthermore, the impact of distance metric selection and confidence aggregation methods generalizes across tasks. For NLP tasks, reversing the distance metric results in an average performance drop of 54.18\%, while reversing the confidence aggregation method leads to an average drop of 57.48\%. For code-related tasks, the average performance drops are 34.13\% and 40.675\%, respectively.}






\begin{table*}[htbp]
    \centering
    \renewcommand{\arraystretch}{1.0}
    \footnotesize
     \resizebox{0.9\columnwidth}{!}{
    \setlength{\tabcolsep}{2pt}
    \begin{tabular}{lcccccccc}
    \toprule
     & AGIEval & TriviaQA  & NQ-open & GSM8K & TruthfulQA-I &  TruthfulQA-T & HumanEval & MBPP \\
    \midrule
    Original & 0.0101 & 0.0027 & 0.018 & 0.0262 & 0.0036 & 0.0232 & 0.0300 & 0.0256 \\ \midrule
    
    \multirow{2}{*}{Reverse DIST} & 0.0197 & 0.0036 & 0.0190 & 0.0399 & 0.0036 & 0.0203 &	0.0778 & 0.0306 \\
        & -94.92\% &	-35.83\% &	-5.84\% &	-52.42\% &	-0.75\% &	+12.78\% &	-158.96\% &	-19.30\% \\

    \midrule
    
    \multirow{2}{*}{Reverse CONF} & 0.0100 & 0.0033 & 0.0162 & 0.0386 & 0.0045 & 0.0192 &	0.0639 & 0.0297 \\
        & +0.91\% &	-23.78\% &	+10.14\% &	-47.70\% &	-25.37\% &	+17.48\% &	-112.58\% &	-15.75\% \\
    
    \hline
    \end{tabular}
     }
    \caption{
        Flipping distance measurement metric (DIST) and confidence (CONF) aggregation methods on Llama-2 and Code Llama
    }\label{tab:diptest_ablation_llama2}
\end{table*}

\begin{table*}[htbp]
    \centering
    \renewcommand{\arraystretch}{1.0}
    \footnotesize
     \resizebox{0.9\columnwidth}{!}{
    \setlength{\tabcolsep}{2pt}
    \begin{tabular}{lcccccccc}
    \toprule
     & AGIEval & TriviaQA  & NQ-open & GSM8K & TruthfulQA-I &  TruthfulQA-T & HumanEval & MBPP \\
    \midrule
    Original & 0.0104 & 0.0024 & 0.0158 & 0.0164 & 0.0034 & 0.0135 & 0.032 & 0.0659 \\ \midrule
    
    \multirow{2}{*}{Reverse DIST} & 0.0234 & 0.0048 & 0.0215 & 0.0203 & 0.0066 & 0.0129 & 0.0297 & 0.1338 \\
        & -125.28\% & -102.22\% & -35.72\% & -23.89\% & -95.38\% & +4.28\% & +7.00\% & -102.99\% \\
        
    \midrule

    \multirow{2}{*}{Reverse CONF} & 0.0126 & 0.0051 & 0.0153 & 0.0166 & 0.0068 & 0.0121 & 0.1771 & 0.0283 \\
        & -21.59\% & -114.56\% & +3.09\% & -0.99\% & -102.87\% & +10.48\% & -168.66\% & +11.35\% \\
        
    \hline
    \end{tabular}
     }
    \caption{
        Flipping distance measurement metric (DIST) and confidence (CONF) aggregation methods on Llama-3
    }\label{tab:diptest_ablation_llama3}
\end{table*}

\begin{table*}[htbp]
    \centering
    
    \renewcommand{\arraystretch}{1.0}
    \footnotesize
     \resizebox{0.9\columnwidth}{!}{
    \setlength{\tabcolsep}{2pt}
    \begin{tabular}{lcccccccc}
    \toprule
     & AGIEval & TriviaQA  & NQ-open & GSM8K & TruthfulQA-I &  TruthfulQA-T & HumanEval & MBPP \\
    \midrule
    Original & 0.0081 & 0.0029 & 0.0178 & 0.0058 & 0.0053 & 0.0039 & 0.0565 & 0.0130 \\ \midrule
    
    \multirow{2}{*}{Reverse DIST} & 0.0382 & 0.0051 & 0.0215 & 0.0079 & 0.0045 & 0.0146 & 0.0322 & 0.0177 \\
        & -373.52\% & -75.78\% & -21.13\% & -36.58\% & +15.74\% & -276.87\% & +42.93\% & -35.92\% \\
        
    \midrule

    \multirow{2}{*}{Reverse CONF} & 0.0344 & 0.0035 & 0.0190 & 0.0067 & 0.0056 & 0.0231 & 0.0817 & 0.0153 \\
        & -326.81\% & -21.92\% & -7.07\% & -16.41\% & -5.75\% & -495.15\% & -44.67\% & -17.77\% \\
        
    \hline
    \end{tabular}
     }
    \caption{
        {Flipping distance measurement metric (DIST) and confidence (CONF) aggregation methods on Qwen2.5}
    }\label{tab:diptest_ablation_qwen}
\end{table*}

\begin{table*}[htbp]
    \centering
    
    \renewcommand{\arraystretch}{1.0}
    \footnotesize
     \resizebox{0.9\columnwidth}{!}{
    \setlength{\tabcolsep}{2pt}
    \begin{tabular}{lcccccccc}
    \toprule
     & AGIEval & TriviaQA  & NQ-open & GSM8K & TruthfulQA-I &  TruthfulQA-T & HumanEval & MBPP \\
    \midrule
    Original & 0.0085 & 0.0032 & 0.0128 & 0.0053 & 0.004 & 0.0059 & 0.0411 & 0.0164 \\ \midrule
    
    \multirow{2}{*}{Reverse DIST} & 0.0071 & 0.0031 & 0.0116 & 0.0051 & 0.0045 & 0.0058 & 0.0438 & 0.0163 \\
        & +16.66\% & +3.28\% & +9.41\% & +3.58\% & -12.35\% & +2.39\% & -6.55\% & +0.73\% \\
        
    \midrule

    \multirow{2}{*}{Reverse CONF} & 0.0093 & 0.0040 & 0.0180 & 0.0093 & 0.0030 & 0.0111 & 0.0320 & 0.0163 \\
        & -9.15\% & -23.63\% & -40.50\% & -74.72\% & +25.25\% & -88.95\% & +22.02\% & +0.66\% \\
        
    \hline
    \end{tabular}
     }
    \caption{
        {Flipping distance measurement metric (DIST) and confidence (CONF) aggregation methods on Phi-4}
    }\label{tab:diptest_ablation_phi}
\end{table*}

\begin{finding}
    \label{finding:4}
    LLMs produce thousands of token-level confidence scores for analysis, making confidence-based performance estimation far more complex than that of single-inference models. {Different strategies for aggregating these confidence indicators can lead to substantial performance variations, with gaps reaching up to 373.52\%.} Our confidence-driven, distribution-aware aggregation method provides a robust foundation for addressing this complexity and serves as an effective starting point for related analyses.
\end{finding}

\section{Discussion} 

\subsection{Algorithm Overhead}
{

There are two major steps in {\ourmethod} that dominate the time complexity of the whole sampling process: (1) Cluster Search in internal state-driven test space partition and (2) intra-cluster sampling. In this section, we briefly analyze both stages' time complexity.

In (1), we perform $t$ times Balanced-Kmeans clustering. The general time complexity of the clustering algorithm ranges from $O(mn^{1.65} )$ to $O(n^3)$~\cite{deMaeyer2023balanced}, where $n$ represents the number of data instances and $m$ denotes the number of iterations. This complexity is influenced by the specific method used to address the balanced constraint. In our study, we employ an auction-based algorithm~\cite{bertsekas1992auction}. Although this approach does not offer the optimal time complexity, it is more amenable to parallelization and can be efficiently executed on GPUs~\cite{gururangan2023scaling}. Notice that in this analysis, we did not consider the layer size (i.e., the number of neurons) because, after dimensionality reduction, the size is significantly reduced to a predefined constant. As such, the upper bound of the time complexity for the search phase is $O(tn^3)$, where $t$ denotes the search budget and is a pre-defined constant in \ourmethod.

In (2), we perform an in-cluster exhaustive search for each sample. Assuming that we have $c$ clusters and $n$ data points with target sample size $n_s$ and there are $n^i_r$ points remaining in the chosen cluster at iteration $i$, then the computation cost is

\begin{equation}
    n^i_r C_{dist} \leq \frac{n}{c} C_{dist}
\end{equation}

where $C_{dist}$ is the cost of the distance function. The condition under which equality holds is that the cluster is being selected for the first time. The cost of both the Kolmogorov–Smirnov test and the one-dimensional Wasserstein distance consists of a sorting operation followed by a linear time computation. By pre-sorting each cluster and maintaining an efficient data structure for the already selected points, the computational cost is primarily determined by $O(i + \frac{n}{c})$. Then, the total cost will be:

\begin{align}
    \sum_{i=1}^{n_s} \frac{n}{c} C_{dist} &= \sum_{i=1}^{n_s} \frac{n}{c} \left[ i + \frac{n}{c} \right] \\
    &= \frac{n}{c}n_s^2 + \frac{n^2}{c^2} n_s
\end{align}

So, the time complexity of this stage is $O(\frac{n}{c}n_s^2 + \frac{n^2}{c^2} n_s)$. While $O(n^3)$ serves as a loose upper bound when $c = 1$ and $n_{s} = n$, this scenario rarely occurs in practice. Our empirical observations indicate that the complexity is typically around $O(n^2)$.

By combining these two stages, we arrive at a time complexity of $O(n^3)$. Although this may appear relatively high, it is considered acceptable for LLM operational testing. Evaluation datasets for LLMs are often limited in size due to the high costs associated with both generation and labeling processes. Typically, the scale of related benchmarks ranges from 100~\cite{chen2021evaluating} to 10,000~\cite{joshi-etal-2017-triviaqa}, making the overhead of our method entirely manageable. Additionally, the most time-consuming stage, clustering, is widely used in the industry and benefits from various frameworks (e.g., NVIDIA cuVS~\cite{nvidia_cuvs}) that support distributed GPU-aided computation. Such optimizations can potentially significantly reduce the actual running time. For more efficient implementation, related stakeholders can consider replacing Balanced-K-means with K-means for lower complexity in (1) and applying better search algorithms (e.g., Metaheuristics, geometric methods, or sampling to find suboptimal data selection choice) instead of the plain exhaustive search in (2). 

We further provide a brief analysis of Confidence-based Stratified Sampling (\csssampling) and Cross Entropy-based Sampling (\cessampling), which frequently emerge as the second-best methods in our evaluation (Table \ref{tab:main_result_llama2} to Table \ref{tab:main_result_phi}). {\csssampling} is relatively straightforward, utilizing output confidence as an indicator and employing a stratified estimator based on variance, with a time complexity of approximately $O(n)$. In contrast, {\cessampling} involves more computationally intensive neuron-based calculations. It seeks to optimize test selection by minimizing the distribution difference between the neuron activation distribution of the last layer in the selected set and that of the entire test set. For LLMs, this requires maintaining a large table of neurons at each iteration, making it significantly slower than {\ourmethod} in our experiments. Consequently, we recommend {\csssampling} when the computational budget is highly constrained.

Based on our overhead analysis, we now discuss a practical approach to integrating active testing methods into a real-world pipeline:
\begin{itemize}[leftmargin=*]
    \item  For a ready-to-release or already deployed LLM-driven system, gather testing data either offline (testing) or online (monitoring). Employ LLM-as-a-judge techniques~\cite{tan2025judgebench} to conduct an unsupervised evaluation of the system under test and assess whether the outcomes align with expectations.
    \item  if any abnormal behaviors are detected, either manually or automatically, initiate active testing on the unlabeled data by selectively choosing data points for labeling. When labeling costs are high and computational resources are plentiful (a common scenario for LLM operations), employ {\ourmethod}. Otherwise, opt for more computationally efficient methods, such as {\csssampling}. This evaluation process assists developers in confirming abnormalities, facilitating root cause analysis, and ultimately enabling the next iteration of model refinement.
\end{itemize}

It is important to note that this is just one use case. {\ourmethod} can be applied in scenarios where more accurate estimation is required, but ground truth labels or model inference costs are prohibitively expensive.
}

\subsection{Possible Improvements for {\ourmethod}} Although {\ourmethod} has shown promising effectiveness across most datasets in our experiments, it still has limitations in accurately estimating certain properties of LLMs, such as truthfulness in specific cases. 

\noindent\textbf{Internal States.} Different properties of tasks can lead to varying behavioral patterns in the internal neurons of LLMs, which in turn influence the estimation accuracy of active testing. To enhance the effectiveness of using internal states as guidance, we propose two promising directions for future research: (1) \textit{Dimension Reduction and Feature Extraction}: The internal states of LLMs are inherently high-dimensional, often comprising 4096 or 8192 dimensions per layer across dozens of layers. Extracting relevant information and reducing dimensionality effectively is critical for improving both efficiency and accuracy. This could involve the intelligent selection of feature layers that contribute most to specific tasks and the development of advanced dimension reduction techniques beyond traditional methods such as PCA. (2) \textit{Engineering Internal States for Task-Specific Properties}: Internal states of LLMs can potentially be engineered to become more responsive to specific properties~\cite{zou2023representation}. A key challenge is designing universal methods to achieve this, thereby enhancing their ability to guide the testing process. Techniques such as linear probing~\cite{marks2024geometry} and representation engineering~\cite{zou2023representation} could play a pivotal role in activating neurons related to specific properties.

{Additionally, our ablation studies in Section~\ref{sec:experiments:rq2} demonstrate that CluSearch is a vital component of {\ourmethod}. However, there is still room for enhancement. Currently, the search budget is fixed, but it could be dynamically adjusted based on the algorithm's feedback. For instance, implementing an early stopping mechanism when the algorithm converges or alerting users if the current budget appears suboptimal could be beneficial. This problem can also be framed as a hyperparameter optimization problem and addressed using algorithms like Bayesian optimization~\cite{snoek2012practical}. Developing more unsupervised metrics beyond inertia to guide the search process is another area for improvement. Given that LLMs may exhibit different behaviors across various tasks, incorporating task-specific metrics could enhance this process. Finally, exploring other search algorithms, such as genetic algorithms, could also be valuable.}


\noindent\textbf{External States.} Our experimental results showed how critical the external aggregation method is with respect to the estimation accuracy of {\ourmethod}. However, this study, as an early exploration work, primarily explores a few straightforward approaches to estimating model confidence based on thousands of output tokens. We believe this area presents numerous opportunities for future exploration within this vast behavioral space. For instance, modeling LLMs as Markov chains~\cite{zekri2024large} could offer a structured way to capture token dependencies. Based on such an interpretation, we can apply more advanced stochastic process analysis methods that might provide deeper insights into the dynamics of LLM outputs.

\subsection{Future Direction for Active Testing}

\noindent\textbf{Dynamic labeling budget adjustment.} While we assumed the labeling budget $n$ was set in advance, our experiments revealed that the optimal sampling size actually varies across different datasets. Within a range, increasing the budget does not always result in lower estimation errors. To illustrate this, consider the following hypothetical scenario: Suppose after sampling $n$ points, we have obtained a perfect estimation $\widetilde{P}$ (i.e., $\frac{1}{n}\sum_{i=1}^{n}p_{i} = \widetilde{P}$), and for the remaining points $p_{j}$ we can not find one that satisfies $p_{j} = \widetilde{P} \times (n+1) - \sum_{i=1}^{n}p_{i}$, then no matter which point we select next, the estimation error will go up. In future work, designing an algorithm that can automatically determine the ideal budget may be beneficial. {One promising approach is to determine when to stop testing based on the observed performance estimation curve during the sampling process. Stakeholders can monitor the convergence of this curve using various time-series analysis techniques. For example, they might perform the slope analysis and compare it to a predefined threshold. Additionally, the accuracy of the performance estimate can be assessed within a statistical framework. For instance, by examining the width of the confidence intervals, one can decide whether the current level of variance is acceptable for drawing reliable conclusions.}

\noindent\textbf{Explore the potential of historical data.} In this work, we assume no prior knowledge of LLMs or their performance. However, in real-world applications, organizations often store historical data on LLMs’ past behavior and performance in their databases. Leveraging this data could significantly enhance the evaluation of LLMs’ capabilities. For instance, performance on one task, such as mathematical problem-solving, might provide useful estimations of LLMs’ code-generation abilities, as both require reasoning skills. Similarly, there is likely a positive correlation between the performance of a base LLM and its fine-tuned versions across various tasks. This opens up opportunities to model the problem from a different perspective: integrating historical performance data to inform and refine evaluation methods.

\noindent {\textbf{More Evaluation Metrics.} Establishing efficient and fair evaluation metrics is crucial for advancing new research directions. In this study, we proposed using the Area Under the Curve (AUC) as a performance indicator. We believe this metric is well-suited for capturing the dynamic performance of various methods, making it particularly appropriate for active testing scenarios. Additionally, AUC serves as a ``smoothed'' version of the plain RMSE. Recall that our study examined evaluation datasets of varying sizes across three magnitudes, employing diverse metrics such as classification accuracy, match rate between answers and ground truth, code pass rate, truthfulness, and informativeness of generated answers. These varied experimental settings cause RMSE at a single sample size to fluctuate significantly—methods that perform best at one point may perform poorly at the next. However, while AUC provides a high-level overview, it lacks finer granularity. For example, how does each method behave when the sample size is rather limited? Future work should consider introducing additional metrics for a more systematic and comprehensive evaluation.}


\section{Related Work}
\label{sec:related}

\subsection{Model Performance Estimation}
Estimating AI models' performance in a label-efficient manner is crucial for applications such as model selection~\cite{zhao2024flasheval}, improvement~\cite{sener2018active}, and performance monitoring~\cite{lu2024characterizing}. 
Most prior research concentrated on classification models, often addressing the estimation problem through learning-based or model-based analysis methods.
For learning-based approaches, a key step is selecting and processing the appropriate features for subsequent learning. 
It is possible to learn the relationship between models' output~\cite{deng2021labels, guillory2021predicting, deng2023confidence, hu2023aries, fu2023estimating} or models' internal states~\cite{miao2023k, miao2024divide} w.r.t their performance. 
{Although these learning-based methods present promising capabilities, they heavily depend on existing labeled data. In contrast, our work aims to establish a general testing framework that does not assume the availability of prior knowledge for training.} 

Model-based analysis approaches offer alternatives by examining models' behaviors, for instance, estimating performance by observing inconsistencies across an ensemble of models~\cite{mehra2024predicting} or analyzing model reactions to input perturbations~\cite{jain2023bring, lee2023unsupervised}. 
Nevertheless, these methods usually focus more on anomaly detection, such as predicting out-of-distribution (OOD) cases, rather than estimating the general performance of models. 

Another widespread evaluation approach involves leveraging LLMs themselves to assess the quality of their outputs~\cite{lin-etal-2022-truthfulqa, lin2023llm, wang2024pandalm, qin2023is, bang-etal-2023-multitask}. 
This solution takes advantage of LLMs' ability on human language understanding and can be scaled up without human intervention. 
For instance, TruthfulQA~\cite{lin-etal-2022-truthfulqa} utilizes a fine-tuned GPT-3 to evaluate truthfulness, while LLM-EVAL~\cite{lin2023llm} introduces a unified schema to label open-domain conversations. 
PandaLM~\cite{wang2024pandalm} adapts a judge language model trained to rank the outputs from other models. 
However, recent studies indicate that LLM-centric evaluations can be biased and may not always provide reliable results, underscoring the necessity for human labeling~\cite{panickssery2024llm}. 
Moreover, the LLM evaluator and active testing are not mutually exclusive, as the latter can be integrated into the evaluation process based on the former to reduce costs further.

\subsection{Active Testing}
Active testing involves sequentially selecting test data for labeling from a pool of unlabeled samples, aiming to estimate a model's behavior under certain metrics. 
{The primary goal is to achieve unbiased sampling and reduce the overall variance of estimations. To accomplish this, related methods leverage either internal or external information for sampling guidance.
Some of the methods utilize internal information of DNN models. For example, CSE \cite{li2019boosting} and PACE~\cite{chen2020practical} both analyze the neuron activation patterns of each test data point to select a representative subset actively. 
Another line of research tries to use the external output information. For example, Kossen \etal~\cite{kossen2021active} explore the connection between active testing and active learning, proposing a framework that selects data points according to acquisition functions. The framework is further refined by ASEs~\cite{kossen2022active}, which introduces a more robust acquisition function with theoretical guarantees. 
However, implementing such functions for testing autoregressive generation models can be challenging, and they often require additional training data for initialization. DiffUse~\cite{ashury2024label} analyzes the models' responses using embedding models and performs corresponding clustering analysis. Finally, DeepSample~\cite{guerriero2024deepsample} is a relatively comprehensive study that compares different internal and external guided testing approaches, but it primarily focuses on pre-LLM DNNs.}



Despite all these advancements, to the best of our knowledge, there is no systematic research on performing effective active testing on LLMs. We hope our work can serve as one of the baselines along the direction and pave the way for more cost-efficient LLM evaluation and development.

\section{Threat to Validity}
\label{sec:validity}

\noindent\textbf{\textit{Threats to Internal Validity}}. One potential threat arises from the randomness introduced by different sampling methods. To address this, we repeat each sampling point 10 times and use the median value for evaluation, which costs in total nearly 1,000 GPU hours and more than 15,000 CPU hours. Other potential threats include the influence of key design choices, such as layer selection and confidence computation methods. To reduce their impact, we conducted systematic studies to assess their effects and provided in-depth discussions to ensure the validity of our conclusions. 

\noindent\textbf{\textit{Threats to External Validity}}. One potential threat lies in the choice of dataset and model, as the results may be biased toward specific selections. To mitigate this, we selected seven datasets across various domains and tested NLP-specific models (Llama-2), code-specific models (Code Llama), and general models (Llama-3, Qwen2.5, Phi-4). These choices represent a diverse range of tasks and capabilities within the limits of our computational budget. In future work, expanding the evaluation to include additional tasks and models would be an interesting direction.

\noindent\textbf{\textit{Threats to Construct Validity}}. A potential threat to construct validity lies in the effectiveness of the evaluation metrics used. To address this, we compute the AUC for the active testing process as a general indicator to compare the performance of different methods in a pool-based setting. For future research, it would be valuable to explore broader sampling proportions and incorporate additional metrics that capture different aspects of related methods.

\section{Conclusion}
\label{sec:conclusion}
In this paper, we introduce a novel active testing framework called {\ourmethod}, designed to select a subset of test data via a multi-stage sampling scheme, thereby accomplishing a comprehensive performance estimation for LLMs.
Different from the existing active testing methods, {\ourmethod} considers the distinct characteristics of LLMs and leverages both internal hidden states and external output confidence traces to collaboratively select a subset of the most representative test data for LLM performance estimation. 
Extensive experiments across a variety of tasks and LLMs have demonstrated the effectiveness of {\ourmethod}, outperforming SOTA baselines on most datasets. Our further in-depth analysis highlights both the challenges and opportunities in developing more advanced adaptive test estimation methods for LLMs, particularly as they serve as general-purpose models across a wide range of tasks. We believe that sample-efficient evaluation is critical to the continuous integration and deployment processes of LLM-driven applications, and our work offers a starting point.
We hope that our exploratory work can inspire further research in this direction, aiming to establish efficient and accurate performance evaluation techniques for LLMs.

\newpage

\bibliographystyle{ACM-Reference-Format}
\bibliography{custom}



\end{document}